\documentclass[prd,amsmath,floats,amssymb,authoryear,floatfix,superscriptaddress,nofootinbib,onecolumn]{revtex4} 

\usepackage{graphicx}
\usepackage{hhline} 
\usepackage{placeins}
\usepackage{caption}
\usepackage{comment}
\usepackage{ulem}
\usepackage{natbib}
\usepackage{multirow}
\usepackage{makecell}
\usepackage{placeins}

\renewcommand{\thetable}{\textbf{\arabic{table}}}

\makeatletter
\renewcommand{\fnum@figure}{\textbf{Figure~\thefigure}}
\renewcommand{\fnum@table}{\textbf{Table~\thetable}}
\makeatother

\usepackage{float}
\usepackage[caption=false]{subfig}

\usepackage[plainpages=false, colorlinks=true, anchorcolor=blue, linkcolor=blue, citecolor=blue, bookmarks=false]{hyperref}

\begin{document}
\preprint{APS/123-QED}

\title{Bayesian and frequentist constraints on dark energy parameterizations using non-CMB data}

\author{Shubham Barua}
 \altaffiliation{Email:ph24resch01006@iith.ac.in}
\author{Ansh Chandrakar}
 \altaffiliation{ph25mscst11009@iith.ac.in}
\author{Shantanu Desai}
 \altaffiliation{Email:shntn05@gmail.com}
\affiliation{
 Department of Physics, IIT Hyderabad Kandi, Telangana 502284,  India}

\begin{abstract}
Data from the Dark Energy Spectroscopic Instrument (DESI) BAO combined with the cosmic microwave background (CMB) have provided evidence for dynamical dark energy (DDE) with a phantom crossing ($w(z) < -1$) when constraining $w_0-w_a$ parameterizations. Given the tension between BAO and CMB data within $\Lambda$CDM, we use only non-CMB data: Type Ia Supernovae (SNe) from the DES-Dovekie compilation, DESI BAO DR2, Gamma-Ray Bursts (GRBs) and Redshift-Space-Distortion (RSD) measurements. We use the dataset combinations GRB$+$BAO, BAO$+$SNe and GRB$+$BAO$+$SNe, each with RSD. We perform both Bayesian and frequentist analyses of two quintessence-inspired thawing dark energy parameterizations: the thawing quintessence parameterization (TQD) and the Algebraic Thawing parameterization (AT), together with the Chevallier-Polarski-Linder (CPL) and the Generalized Emergent Dark Energy parameterizations (GEDE). The GRB sample, used in this work, contributes only marginally to the constraints, while SNe tighten them considerably. The CPL equation of state shows a phantom crossing at $z\sim 0.5$, while TQD and AT improve the best-fit $\chi^2$ over $\Lambda$CDM, without showing phantom crossing. When SNe are included, TQD is preferred by the Bayes factor (moderately), AIC and DIC (positively). For CPL and AT, AIC and DIC prefer the extended model positively, while the Bayes factor weakly prefers $\Lambda$CDM over CPL and is inconclusive for AT. This difference is consistent with the Occam penalty from the prior volume in the Bayes factor. GEDE, with the DE density restricted to be suppressed at high redshifts is disfavored by all three metrics. Hence, the preference for DDE depends on the parameterization, the dataset and the comparison metric used. 
\end{abstract}


\maketitle

\section{Introduction}

Over the years, the standard $\Lambda$CDM cosmological model has emerged as the concordance model of cosmology, successfully describing observations ranging from the anisotropies of the Cosmic Microwave Background (CMB) to the formation and evolution of large-scale structures. Despite its remarkable success, increasing observational precision has revealed several tensions within this framework, most notably the Hubble tension - a statistically significant disagreement between the CMB-inferred value of the Hubble constant ($H_0 = 67.19\pm0.38$km/s/Mpc)~\cite{camphuis_2026} and the locally measured value from the distance ladder ($H_0 = 73.50\pm0.81$ km/s/Mpc)~\cite{h0dn_2026}, amounting to a discrepancy of $7.1\sigma$. This persistent discrepancy has motivated an extensive search for both extensions to and alternatives beyond the standard $\Lambda$CDM framework~\cite{cosmoverse}.

Beyond the Hubble tension, the physical nature of dark energy (DE) remains one of the central open questions in cosmology. Within $\Lambda$CDM, the cosmic acceleration which was discovered using Type Ia supernovae (SNe)~\cite{perlmutter_1999, adam_1998}, is usually attributed to a cosmological constant $\Lambda$ with equation of state (EOS) $w(z)=-1$. Recent baryon acoustic oscillation (BAO) measurements from the second data release (DR2) of the Dark Energy Spectroscopic Instrument (DESI)~\cite{desi_2025}, however, have reported evidence for dynamical dark energy (DDE) when the data are analyzed using the Chevallier-Polarski-Linder (CPL) parameterization. Depending on the combination of datasets employed, the preference for DDE reaches $3.1 \sigma$ for CMB$+$BAO and $2.8-4.2 \sigma$ for CMB$+$BAO$+$SNe, depending on the SNe dataset employed. The EOS exhibits a transition into the phantom regime ($w(z)<-1$) at $z \sim 0.4$, with $w(z)<-1$ at higher redshifts. 

The possibility that the accelerated expansion of the universe is driven by a dynamical component rather than a cosmological constant has motivated a broad range of theoretical models. These include canonical quintessence scalar field theories~\cite{ratra_1988, peebles_1988, wetterich_1988, joshua_1995, caldwell_1998, turner_1997, coble_1997}, non-canonical "k-essence" theories with tracking behavior~\cite{picon_2000, picon_2001}, holographic DE models~\cite{li_2004}, modified gravity scenarios~\cite{saridakis_2021}, and extensions of the standard cosmological framework~\cite{lodha_2025}. Theory-specific analyses have extensively confronted such models with cosmological observations~\cite{bozek_2008, akrami_2019, sukannya_2024, sukannya_2025, raveri_2019, george_2024, ruchika_2023, gupta_2012, dutta_2007, nils_2023, ramadan_2024}. Given the diversity of theoretical possibilities, phenomenological parameterizations of $w(z)$ provide a complementary approach for constraining departures from $\Lambda$CDM without committing to a particular underlying theory. Different parameterizations can capture different classes of DE dynamics: some are tailored to specific theoretical scenarios, while more general forms can encompass a broader range of models.

Among these parameterizations, the CPL form, $w(a) = w_0 + w_a(1-a)$, has become one of the most widely used due to its simplicity and ability to capture a broad range of departures from $\Lambda$CDM. This parameterization was shown to capture and reproduce the dynamics of freezing and thawing scalar-field DE models~\cite{linder_2008, linder_2024}. Thus, within the regime of simple freezing and thawing single-field models, CPL can provide a useful phenomenological description of their background dynamics. However, its two-parameter form is not designed to capture the full range of possible DE dynamics, particularly those arising from more general theoretical constructions~\cite{savvas_2025, gabriele_2026, akarsu_2026}. More specifically, CPL is a phenomenological parameterization rather than a description derived from a specific microphysical theory. 

An additional complication arises when interpreting constraints in the CPL parameter space. Cosmological observables depend on $w(z)$ primarily through the expansion history and its integrals rather than directly on $w(z)$. Consequently, different functional forms of $w(z)$ can produce very similar expansion histories and observational signatures despite differing substantially in their detailed evolution~\cite{carlos_2020, putter_2008, linder_2003}. In particular, it has been shown that several classes of quintessence models map into the same region of parameter space currently favored by observations, even though these models never cross the phantom boundary by construction~\cite{shlivko_2024}. Therefore, a preference for phantom-crossing in the CPL parameter space does not necessarily imply the presence of a phantom DE component, but may instead indicate a more general preference for DDE.

This issue is particularly relevant for thawing quintessence scenarios, in which a canonical scalar field remains frozen near $w(z) = -1$ at early times, due to Hubble friction, before evolving away from the cosmological constant solution at late times. Recent analyses have found that thawing quintessence models provide competitive fits to cosmological expansion relative to $\Lambda$CDM~\cite{park_2026, souza_2025, akrami_2025, alestas_2026, guillaume_2025}. The preference becomes stronger when the restriction to canonical quintessence is relaxed and the EOS is allowed to enter the phantom regime. Phantom-like behavior can arise in theories with non-canonical kinetic terms~\cite{chen_2026, fotis_2026, goldstein_2026, ren_2026}, non-minimal gravitational couplings~\cite{wolf_2025, brax_2025, adam_2026, lopez_2025, nojiri_2025, nojiri_2026, ye_2025}, non-standard dark matter evolution~\cite{chen_2025, utkarsh_2026, giani_2025, yang_2025} and interacting dark sectors~\cite{toomey_2025, alek_2025, justin_2025, andriot_2025}.

The DESI-preferred CPL solutions cross into the phantom regime at $z \sim 0.4$, corresponding to $w(z) < -1$ at higher redshifts. For a DE component with positive energy density, this regime violates the null energy condition and therefore also the weak energy condition~\cite{carroll_2003}. Such behavior is excluded for a minimally coupled canonical scalar field, and the corresponding CPL solutions therefore cannot be interpreted directly as canonical quintessence histories. Nevertheless, as discussed above, the mapping between the underlying EOS and the CPL parameters is not unique. The CPL region favored by observations can therefore provide an indication of departures from $\Lambda$CDM without uniquely identifying the underlying physical realization of DE.

The distinction between generic and theory-specific parameterizations has motivated recent efforts to construct and compare parameterizations tailored to particular classes of DE models. In particular, \cite{shlivko_2025} focused on thawing DE theories, for which $w(z) \rightarrow -1$ at early times, and compared several thawing-specific parameterizations with more general CPL-like forms according to their ability to reproduce realistic thawing DE dynamics. Such comparisons are important because increasing the flexibility of a phenomenological EOS through additional degrees of freedom~\cite{ewan_2013, pang_2025, joao_2025, escamilla_2025, saddam_2026, mehdi_2024} can come at the cost of larger uncertainties and stronger parameter degeneracies~\cite{huterer_2005}.

Recent work has demonstrated that the preference for phantom crossing over quintessence is driven primarily by CMB and BAO data due to an improvement in the consistency between them~\cite{shlivko_2026}. Within $\Lambda$CDM, these datasets prefer slightly different values of the present matter density parameter, with $\Omega_m^{\mathrm{BAO}} < \Omega_m^{\mathrm{CMB}}$ (see also, \cite{colgain2_2025, zhou_2026}). This discrepancy becomes more pronounced when the analysis is restricted to quintessence models, whereas allowing a phantom-crossing EOS improves the consistency between the datasets. 

These considerations point to two tensions involving CMB data: the Hubble tension and the offset between the CMB and BAO values of $\Omega_m$ within $\Lambda$CDM. Since the preference for DDE may be affected by these tensions, it is useful to examine the same parameterizations without CMB data. Motivated by this, in the present work we investigate five DE parameterizations using a combination of non-CMB probes namely, BAO~\cite{desi_2025}, SNe~\cite{des_2026}, gamma-ray bursts (GRBs)~\cite{khadka_2021} and Redshift-space distortions (RSD)~\cite{mitra_2026}. The DE frameworks are the standard $\Lambda$CDM model, the CPL parameterization~\cite{linder_2003, chevallier_2001}, the thawing quintessence parameterization (TQD)~\cite{copeland_1998}, the Algebraic Thawing (AT) parameterization~\cite{linder_2008} and the generalized emergent dark energy (GEDE) parameterization~\cite{li_2020}. We derive constraints on the parameters of each framework and assess their relative performance with respect to $\Lambda$CDM. The model comparison is based on the Bayes factor and the Deviance Information Criterion (DIC) from the Bayesian runs, and on the Akaike Information Criterion (AIC) from the frequentist best-fit runs.

This work is organized as follows: In Section~\ref{sec2}, we present a brief description of the models considered and the methodology and datasets is discussed in Section~\ref{sec3}. In Section~\ref{sec4}, we present and discuss our results and we conclude in Section~\ref{sec5}.

\section{Model framework}
\label{sec2}

Here we describe the models analyzed in our work. We consider the Friedmann-Lemaitre-Robertson-Walker (FLRW) metric. Solving the Friedmann equations using the FLRW metric, we get the expansion history of the universe
\begin{equation}
\label{eqn1}
    H(z)^2 = H_0^2\left[\Omega_{m0}(1+z)^3 + \Omega_k(1+z)^2 + f_{DE}(z)\right],
\end{equation}
where $\Omega_k$ is the curvature of the universe, which we assume to be 0 in this work and $f_{DE}(z)$ is the term representing the DE dynamics
\begin{equation}
\label{eqn2}
    f_{DE}(z) = \Omega_{DE,0}\exp\left[3\int_0^z\frac{1+w(z')}{1+z'}dz'\right],
\end{equation}
where $\Omega_{DE,0} = 1 - \Omega_{m0} - \Omega_k$. $\Omega_{m0}$ is the current dimensionless matter density of the universe.

We consider the following DE parameterizations:

\begin{itemize}
    \item \textbf{$\Lambda$CDM} - This is the usual standard model of cosmology where $w(z)=-1$.
    \item \textbf{CPL} - This is the widely used phenomenological parameterization~\cite{linder_2003, chevallier_2001} given by the EOS 
    \begin{equation}
    \label{eqn3}
    w(z) = w_0+w_a\frac{z}{(1+z)}.
    \end{equation} 
    $w_0$ is the present-day value of the DE EOS and $w_a$ characterizes the redshift evolution of the EOS. Extrapolating the CPL framework to very high redshifts ($a \rightarrow 0$) raises theoretical issues. Such behavior might be inconsistent with physically-motivated scalar field DE models. Hence, it should not be interpreted as a fundamental DE model.
    \item \textbf{TQD} - The EOS for the thawing quintessence parameterization~\cite{copeland_1998, souza_2025, rodrigues_2025} is given by 
    \begin{equation}
    \label{eqn4}
    w(z)=-1+\delta\left(\frac{1}{1+z}\right)^{\gamma}.
    \end{equation} 
    Quintessence models of DE predict an EOS satisfying $w(z) > -1$. To generalize a class of quintessence potentials, Ref.~\cite{carvalho_2006} proposed the ansatz, $1/\rho_{\phi}(d\rho_{\phi}/da) = -3\delta/a^{1+\gamma}$ which when combined with the continuity equation, $\dot\rho_{\phi}+3H\rho_{\phi}[1+w(a)]=0$ gives the EOS Eqn.~\ref{eqn4}. $\delta$ sets the present-day EOS value and $\gamma$ controls the rate of deviation from $w(z)=-1$. When both $\delta >0$ and $\gamma>0$, the model reproduces the dynamics of thawing quintessence, while in general, it captures both thawing and freezing behaviors of scalar field models.
    \item \textbf{AT} - We consider the Algebraic Thawing parameterization~\cite{shlivko_2025, linder_2008, linder_2015} given by
    \begin{equation}
    \label{eqn5}
        w(z) = -1 + (1+w_0)a^p\left(\frac{1+b}{1+ba^{-3}}\right)^{1-p/3},
    \end{equation}
    where $a=(1+z)^{-1}$ and $w_0$ represents the present-day EOS value. Following~\cite{linder_2015}, we consider $b=0.5$. We let the parameter $p=1$ effectively making Eqn.~\ref{eqn5} a one-parameter DE parameterization.
    \item \textbf{GEDE} - This parameterization~\cite{li_2020, yang_2021, yang_2026} is a phenomenological model where DE EOS increases from $w(z) < -1$ at high redshifts and becomes $w(z) = -1$ at later times. The EOS is given by
    \begin{equation}
    \label{eqn6}
        w(z) = -1 - \frac{\beta}{3}\left(1-\tanh \left[\beta\ln\left(\frac{1+z_t}{1+z}\right)\right]\right),
    \end{equation}
    where $\beta$ is the transition width and the transition redshift is determined by $\Omega_{DE}(z_t) = \Omega_{m0}(z_t)$. $\beta$ is the only free parameter here. For $z \gg z_t$, $w(z) \rightarrow -1 -2\beta/3$ and for $z \ll z_t$, $w(z) \rightarrow -1$. The prior on $\beta$ is chosen to be $[0, 4.34]$ (same as in Ref.~\cite{yang_2026}) in order to suppress the DE density compared to $\Lambda$CDM at high-redshifts.
\end{itemize}

\section{Datasets and methodology}
\label{sec3}

The datasets utilized in this work are as follows:
\begin{itemize}
    \item \textbf{BAO} - We use DESI DR2 BAO measurements~\cite{desi_2025} from multiple tracers in seven different redshift bins. These consist of the spherically averaged distance ($D_V$), the Hubble distance ($D_H$) and the comoving angular diameter distance ($D_M$) normalized to $r_d$ and the correlations between them.
    \item \textbf{SNe} - We use the recent DES-Dovekie SNe compilation (recalibrated DESY5 sample) which consists of 1623 DES SNe and 197 low-$z$ SNe~\cite{des_2025, des_2026}.
    \item \textbf{GRBs} - We use the A118 GRB sample compilation from~\cite{khadka_2021, cao_2023} which consists of 118 long GRBs spanning a redshift range of $0.3399\leq z\leq8.2$. The isotropic radiated energy $E_\mathrm{iso}$ of a GRB in its rest frame is given by $E_\mathrm{iso}= 4\pi D_L^2S_\mathrm{bolo}/(1+z)$, where $D_L$ is the luminosity distance and $S_\mathrm{bolo}$ is the bolometric fluence computed in the standard rest-frame energy band $1-10^4$ keV. The peak energy of a GRB source is $E_p=(1+z)E_{p,\mathrm{obs}}$, where $E_{p,\mathrm{obs}}$ is the observed peak energy. The Amati correlation~\cite{amati_2008, amati_2009} relates $E_\mathrm{iso}$ and $E_p$ through the relation $\log E_\mathrm{iso}=x+y \log E_p$ \footnote{We use $x$ and $y$ instead of the $a$ and $b$ used in Ref.~\cite{khadka_2021} to avoid confusion with the scale factor $a$ and the AT parameter $b$.}. As shown in Ref.~\cite{khadka_2021}, the A118 GRB sample is standardizable and suitable for use in cosmological studies because the Amati correlation parameters $x$ and $y$ are cosmological-model independent. 
    \item \textbf{RSD} - We use seven growth rate measurement $f\sigma_8(z)$ datapoints (Table~\ref{table2}) compiled from SDSS MGS~\cite{howlett_2015}, BOSS DR12~\cite{alam_2017} and the final eBOSS~\cite{alam_2021} galaxy and quasar surveys. This compilation was also used in~\cite{mitra_2026} and we follow their procedure in order to compute the growth observables. The equation for the subhorizon linear growth factor $D(a)$ is given by:
    \begin{equation}
    \label{eqn7}
        D'' + (2 + \ln(H)')D'-1.5\Omega_m(x)D=0,
    \end{equation}
    where $'$ denotes derivative with respect to $x$ and $x\equiv-\ln(1+z)=\ln(a)$. The matter density fraction is $\Omega_m(x) = \Omega_{m0}\exp(-3x)/E^2(x)$ ($E(x)=H(x)/H_0$). This second-order ODE is solved using the fourth-order Runge Kutta solver integrated from $x = -4$ (corresponding to $z\sim53.6$ which is deep in the matter dominated epoch where $D(a)\propto a$) to the present day $x=0$ ($z=0$). The initial conditions at $x_\mathrm{ini}=-4$ are given by $D(x_\mathrm{ini})=\exp(x_\mathrm{ini})$ and $D'(x_\mathrm{ini})=\exp(x_\mathrm{ini})$. These conditions are consistent with the standard matter dominated growing mode attractor solution $D(a)\propto a$. The growth rate (defined as $f(z)\equiv d \ln D/d \ln a$) allows us to compute the growth observable: 
    \begin{equation}
    \label{eqn8}
        f\sigma_8(z)=\sigma_8(z=0)f(z)\frac{D(z)}{D(0)}.
    \end{equation}
    Henceforth, we will be using $\sigma_8$ to represent $\sigma_8(z=0)$ unless stated otherwise.
\end{itemize}

In order to perform our analysis, we consider the GRB$+$BAO (GD), BAO$+$SNe (DS) and GRB$+$BAO$+$SNe(GDS) dataset combinations. In each combination, we include RSD measurements as well. Since BAO observations constrain the late-time expansion history through the ratio of the cosmological distances to $r_d$, they exhibit a degeneracy between $H_0$ and $r_d$. As the BAO dataset is included in all the  combinations we consider in this work, we therefore present constraints on the parameter combination $H_0r_d$ rather than $H_0$ alone, to avoid imposing a specific calibration of the absolute cosmological scale. It is worth noting that even when adopting the standard $\Lambda$CDM model and imposing a calibration prior of $r_d \in \mathcal{N}(147.09, 0.26)$ Mpc~\cite{planck_2020}, the results presented in this work remain essentially unchanged . This is expected, as the physics governing the pre-recombination sound horizon does not directly affect the late-time dynamics of DE.

We perform the Bayesian analysis using the Importance Nested Sampling sampler \texttt{Nautilus}~\cite{nautilus}. We use 5000 live points and set the target effective sample size to 50000.

For the frequentist analysis, we use the \texttt{Minuit} minimizer~\cite{minuit} from the \texttt{iminuit}\footnote{\url{https://zenodo.org/records/17565861}} package. For a parameter of interest $\theta$, we construct the profile likelihood (PL) by fixing $\theta$ to a grid of values and minimizing $\chi^2$ over all remaining model parameters at each grid point. The resulting profile $\Delta \chi^2(\theta) = \chi^2(\theta) - \chi^2_\mathrm{min}$ is then plotted as a function of $\theta$. The $1\sigma$ confidence interval is obtained from the points at which the profile crosses $\Delta \chi^2 = 1$. Here, $\chi^2_\mathrm{min}$ denotes the global minimum obtained by minimizing $\chi^2$ over all model parameters simultaneously, while $\chi^2(\theta)$ denotes the minimum $\chi^2$ at each fixed value of $\theta$ when profiling over all remaining parameters. 

For a one-sided upper/lower limit, we use $\Delta \chi^2 = 2.71$, corresponding to the usual asymptotic $95\%$ one-sided confidence limit for one parameter of interest. We note that this threshold is based on the asymptotic likelihood-ratio approximation and may not provide exact frequentist coverage when the parameter is close to a physical boundary or when the regularity conditions underlying the asymptotic result are not satisfied~\cite{herold_2025, colgain_2025, barua_2026}. 

To compare the different parameterizations, we compute the logarithm of the Bayes factor as the difference between their log Bayesian evidences, $\ln B_{1i} = \ln Z_1 - \ln Z_i$, where $\ln Z_i$ denotes the log Bayesian evidence for parameterization $i$. We take the $\Lambda$CDM model as the reference model, such that $\ln Z_1$ corresponds to its log Bayesian evidence. Positive values of $\ln B_{1i}$ indicate a preference for $\Lambda$CDM while negative values favor the alternative DE parameterization. The strength of evidence, quantified by $|\ln B_{1i}|$, is classified as \textit{inconclusive} for $|\ln B_{1i}|<1$, \textit{weak} for $1.0 \leq |\ln B_{1i}| < 2.5$, \textit{moderate} for $2.5 \leq |\ln B_{1i}| < 5.0$ and \textit{strong}  $|\ln B_{1i}| \geq 5.0$~\cite{Trotta08}.

We also compute AIC and DIC~\cite{krishak_2020, liddle_2007, marek_2007, kunz_2006, ratra_2024}. The AIC is defined as:

\begin{equation}
\label{eqn9}
    \mathrm{AIC} = \chi^2_\mathrm{min} + 2n, 
\end{equation}
where $n$ is the number of free model parameters. DIC is a Bayesian model selection criterion defined as
\begin{equation}
\label{eqn10}
    \mathrm{DIC} = \chi^2(\bar{\theta}) + 2p_D,
\end{equation}
where $p_D = \overline{\chi^2} - \chi^2(\bar{\theta})$. Here, $\overline{\chi^2}$ is the average of the $\chi^2$ values over the MCMC samples, and $\bar{\theta}$ is the vector of posterior mean values of the model parameters. Thus, $\chi^2(\bar{\theta})$ is the $\chi^2$ evaluated at the posterior mean parameter values. 

When comparing models, we consider $\Lambda$CDM as the reference model and define:
\begin{align}
    \Delta\mathrm{AIC} &= \chi^2_\mathrm{min}(\mathcal{M}) - \chi^2_\mathrm{min}(\Lambda \mathrm{CDM})+ 2k \label{eqn11} \\
    \Delta\mathrm{DIC} &= \mathrm{DIC}(\mathcal{M}) - \mathrm{DIC}(\Lambda\mathrm{CDM}), \label{eqn12}
\end{align}
where $k$ is the number of additional free parameters in $\mathcal{M}$ compared to $\Lambda$CDM and $\mathcal{M}$ is the model under consideration, namely CPL, TQD, GEDE or AT. Following the standard interpretation~\cite{liddle_2007}, the strength of evidence is given by: \textit{weak} for $0\leq|\Delta\mathrm{AIC}, \Delta\mathrm{DIC}|<2$, \textit{positive} for $2\leq|\Delta\mathrm{AIC}, \Delta\mathrm{DIC}|<6$, \textit{strong} for $6\leq|\Delta\mathrm{AIC}, \Delta\mathrm{DIC}|<10$, and \textit{very strong} for $|\Delta\mathrm{AIC}, \Delta\mathrm{DIC}|\geq 10$. Negative values of $\Delta\mathrm{AIC}(\Delta\mathrm{DIC})$ favor the parameterization under consideration, while positive values favor $\Lambda$CDM.

The Bayesian priors are listed in Table~\ref{table1}.

\begin{table}[ht]
\caption{Parameter priors}
\label{table1}
\centering
    \begin{tabular}{c|c}
    \hline
    \thead{Parameters} & \thead{Priors} \\
    \hline
    \hline
    $H_0r_d$ [km/s]  & $\mathcal{U}[7000,13000]$ \\
    $\Omega_{m0}$ & $\mathcal{U}[0, 1]$ \\
    $\sigma_8$ & $\mathcal{U}[0.4, 1.2]$ \\
    $w_0$      & $\mathcal{U}[-3, 2]$ \\
    $w_a$      & $\mathcal{U}[-3, 1]$ \\
    $\delta$   & $\mathcal{U}[0, 1]$ \\
    $\gamma$   & $\mathcal{U}[0, 15]$ \\
    $\beta$    & $\mathcal{U}[0, 4.34]$ \\
    $x$  & $\mathcal{U}[0, 300]$ \\
    $y$ & $\mathcal{U}[0, 5]$ \\
    $\sigma_\mathrm{int}$ & $\mathcal{U}[0, 5]$ \\
    \hline
    \end{tabular}
\end{table}

\begin{table}[ht]
\caption{$f\sigma_8$ measurements.}
\label{table2}
\centering
    \begin{tabular}{c|c}
    \hline
    \thead{$z$} & \thead{$f\sigma_8$} \\
    \hline
    $0.15$ & $0.530\pm0.160$ \\
    $0.38$ & $0.440\pm0.040$ \\
    $0.51$ & $0.458\pm0.038$ \\
    $0.698$ & $0.473\pm0.033$ \\
    $0.85$ & $0.315\pm0.095$ \\
    $1.48$ & $0.462\pm0.045$ \\
    $2.33$ & $0.387\pm0.081$ \\
    \hline
    \end{tabular}
\end{table}


\section{Results and discussion}
\label{sec4}

\begin{figure}[ht]
    \centering
    \subfloat[Bayesian constraints on $\Lambda$CDM \label{1a}]{\includegraphics[width=0.33\textwidth,keepaspectratio]{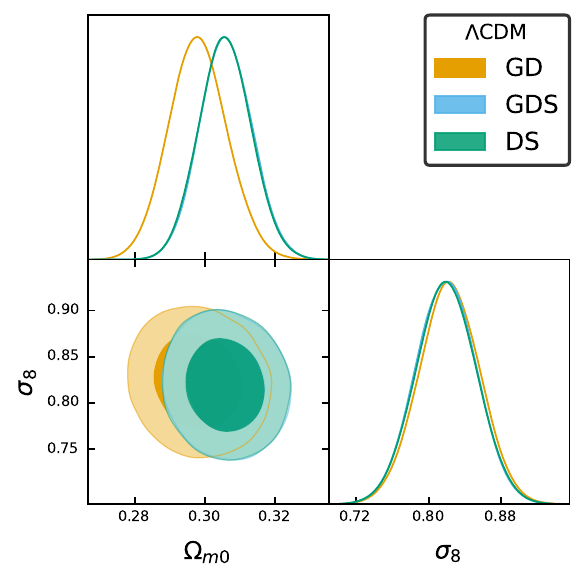}}
    \subfloat[$\Omega_{m0}$\label{1b}]{\includegraphics[width=0.33\textwidth,keepaspectratio]{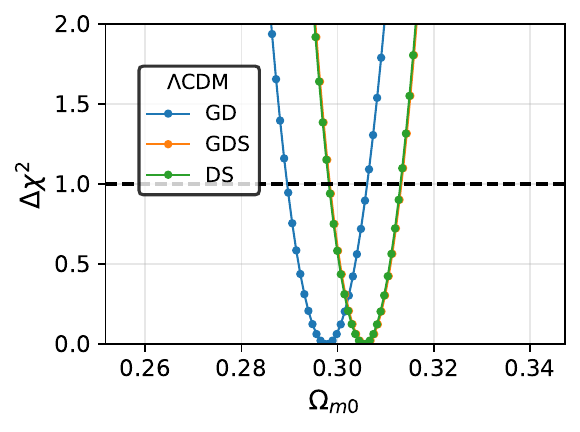}}
    \subfloat[$\sigma_8$\label{1c}]{\includegraphics[width=0.33\textwidth,keepaspectratio]{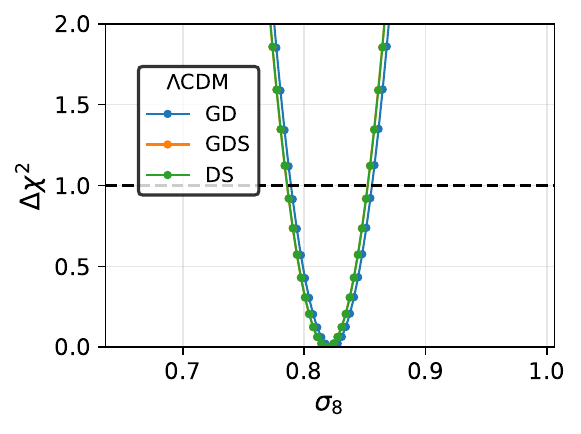}}
    \caption{Bayesian and frequentist constraints on the $\Lambda$CDM model. Top row: Bayesian posterior contours (Fig.~\ref{1a}); contours represent 68\% and 95\% credible regions. Bottom row: PL constraints on $\Omega_{m0}$ (Fig.~\ref{1b}) and $\sigma_8$ (Fig.~\ref{1c}). The $\Delta \chi^2 = 1$ line marks the 68\% confidence interval.}
    \label{fig1}
\end{figure}

\begin{figure}[ht]
    \centering
    \subfloat[Bayesian constraints on CPL \label{2a}]{\includegraphics[width=0.6\textwidth,keepaspectratio]{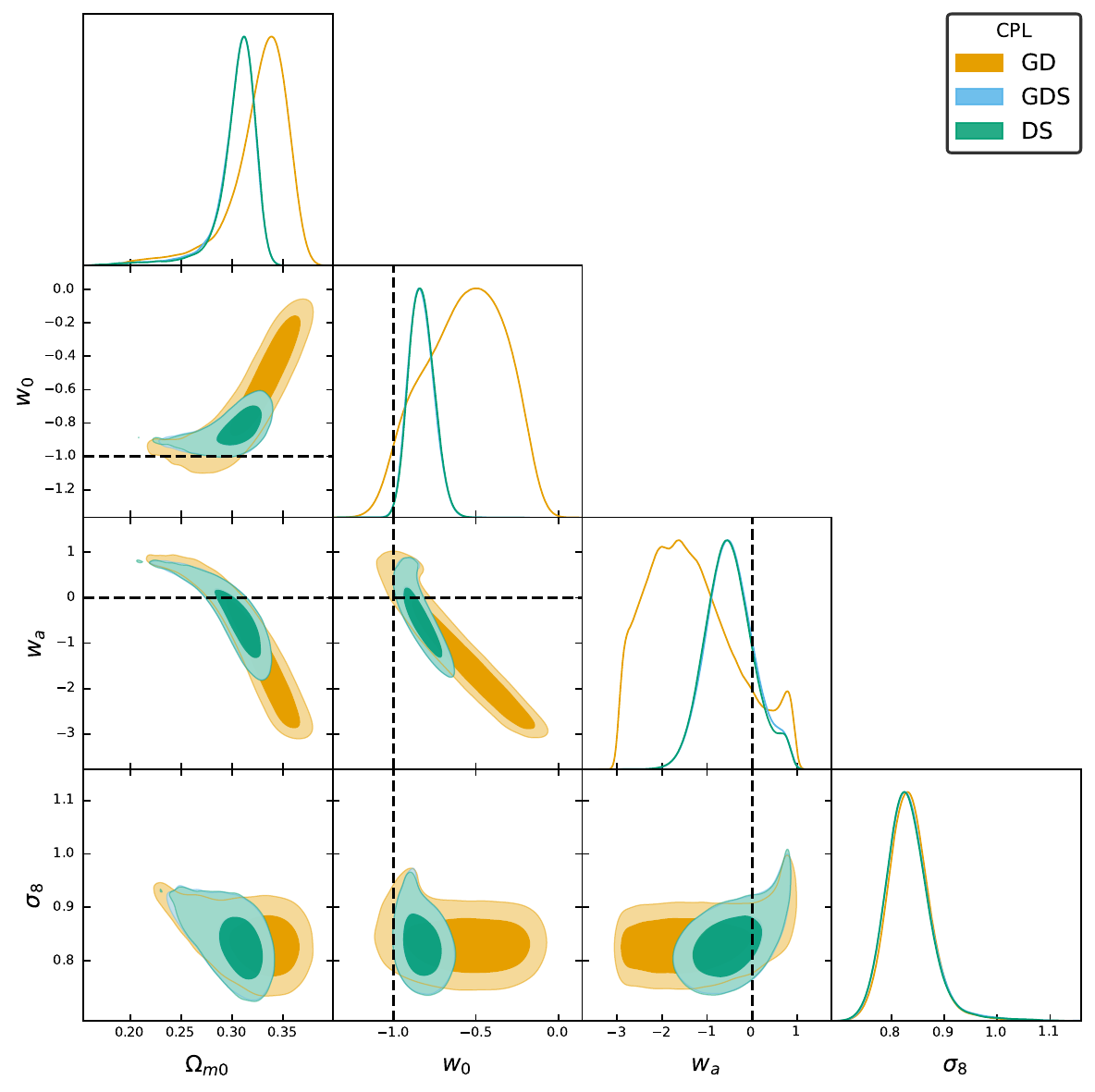}}
    \subfloat[EOS \label{2b}]{\includegraphics[width=0.4\textwidth,keepaspectratio]{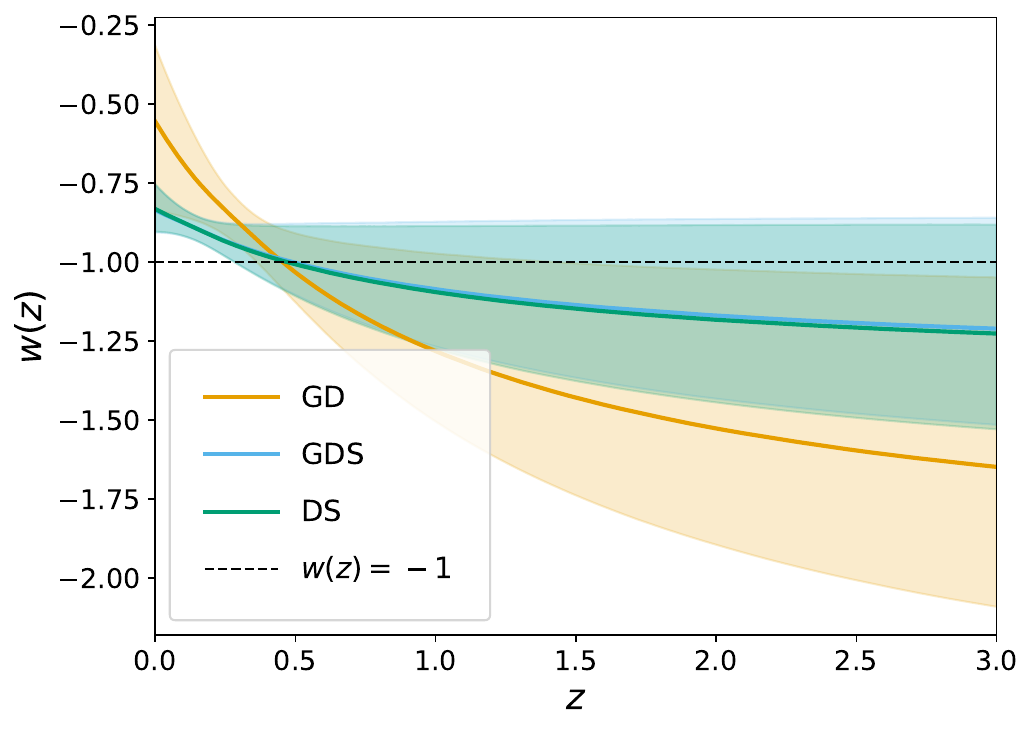}}\\
    \subfloat[$\Omega_{m0}$\label{2c}]
    {\includegraphics[width=0.25\textwidth,keepaspectratio]{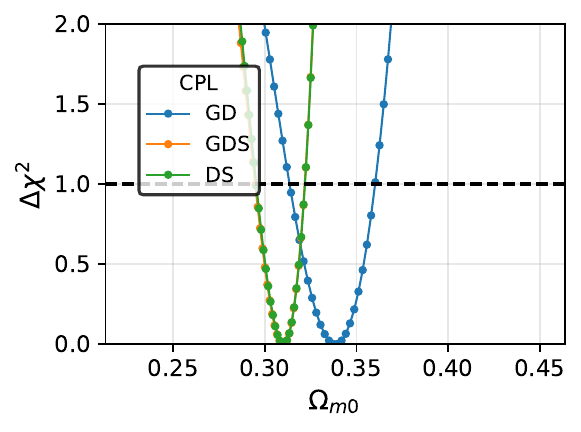}}
    \subfloat[$w_0$\label{2d}]{\includegraphics[width=0.25\textwidth,keepaspectratio]{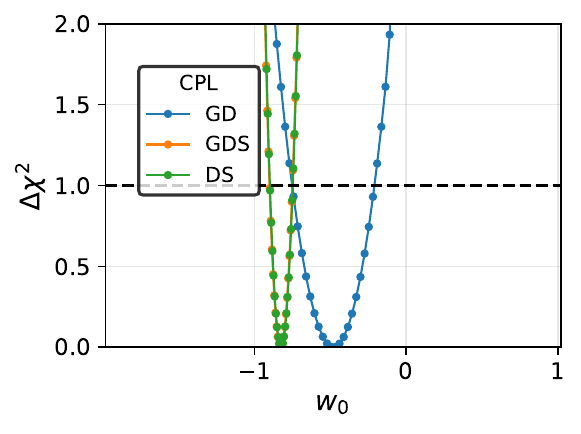}}
    \subfloat[$w_a$\label{2e}]{\includegraphics[width=0.25\textwidth,keepaspectratio]{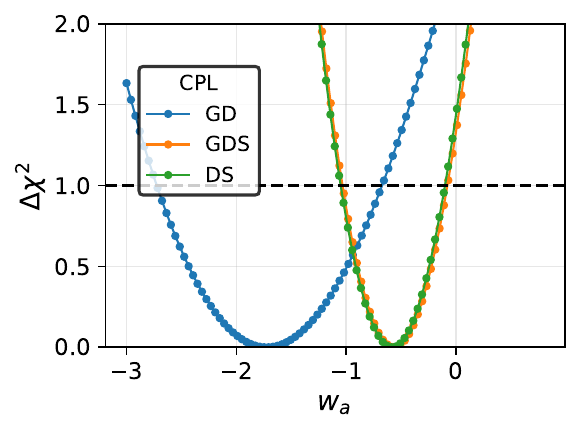}}
    \subfloat[$\sigma_8$\label{2f}]{\includegraphics[width=0.25\textwidth,keepaspectratio]{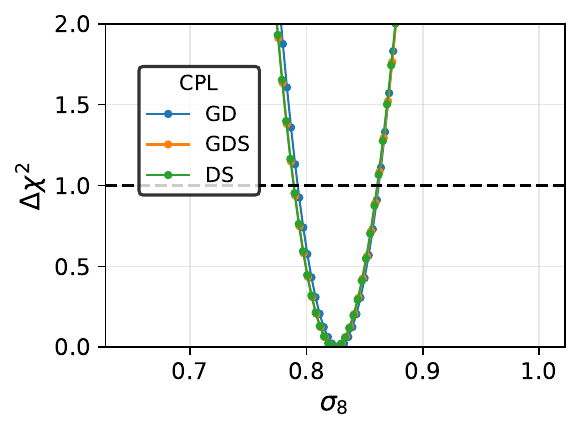}}
    \caption{Bayesian and frequentist constraints on the CPL model. Top row: Bayesian posterior contours (Fig.~\ref{2a}) and the corresponding EOS (Fig.~\ref{2b}); contours represent 68\% and 95\% credible regions. Bottom row: PL constraints on $\Omega_{m0}$ (Fig.~\ref{2c}), $w_0$ (Fig.~\ref{2d}), $w_a$ (Fig.~\ref{2e}) and $\sigma_8$ (Fig.~\ref{2f}). The $\Delta \chi^2 = 1$ line marks the 68\% confidence interval.}
    \label{fig2}
\end{figure}

\begin{figure}[ht]
    \centering
    \subfloat[Bayesian constraints on TQD \label{3a}]{\includegraphics[width=0.6\textwidth,keepaspectratio]{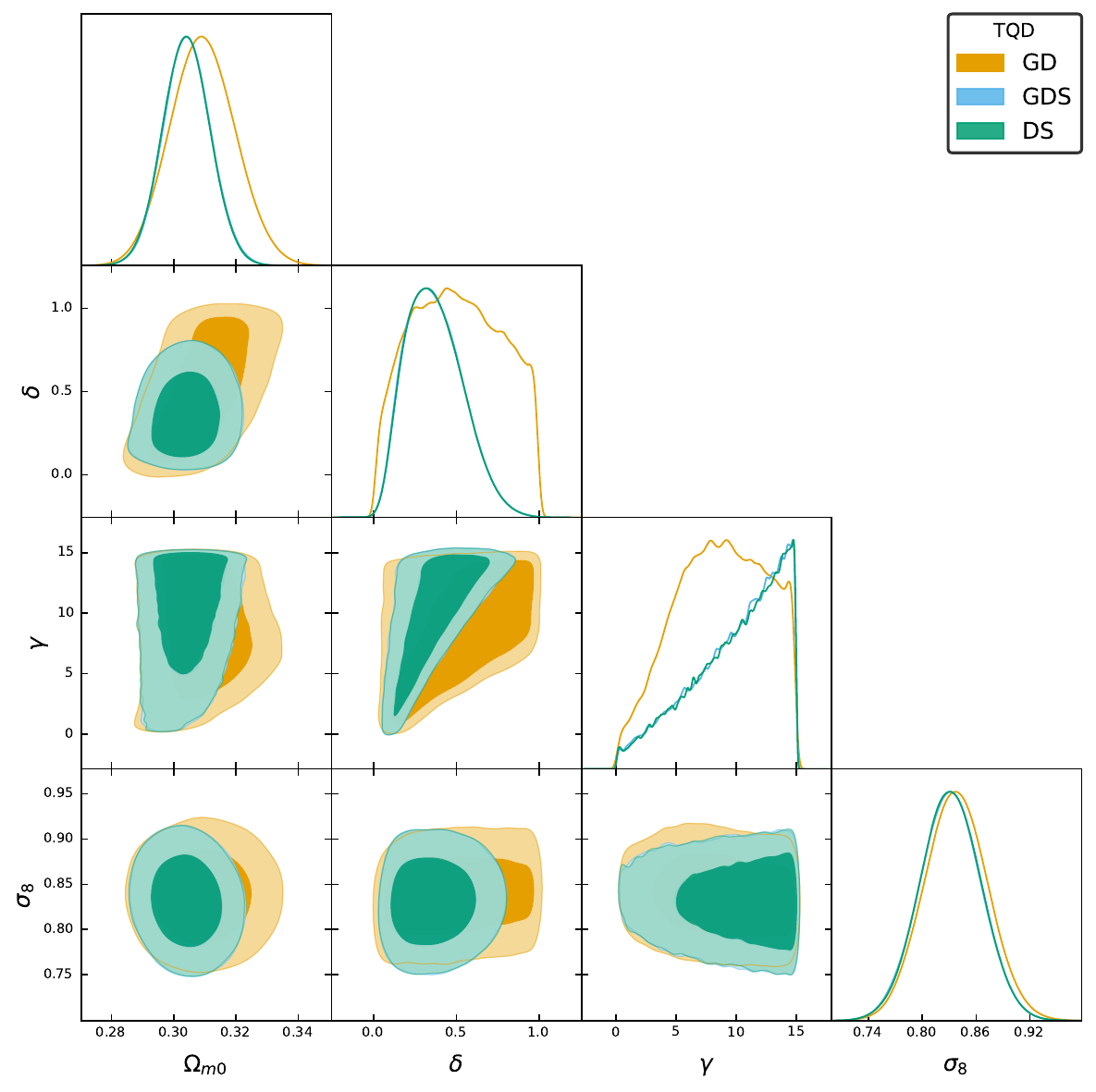}}
    \subfloat[EOS \label{3b}]{\includegraphics[width=0.4\textwidth,keepaspectratio]{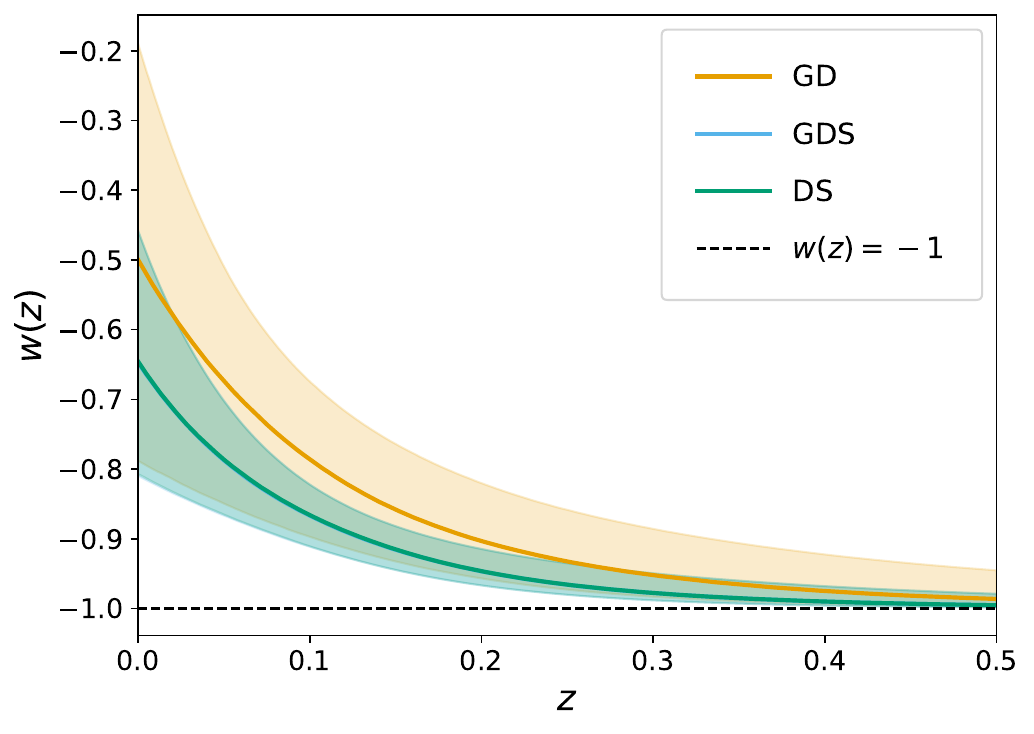}}\\
    \subfloat[$\Omega_{m0}$\label{3c}]
    {\includegraphics[width=0.25 \textwidth,keepaspectratio]{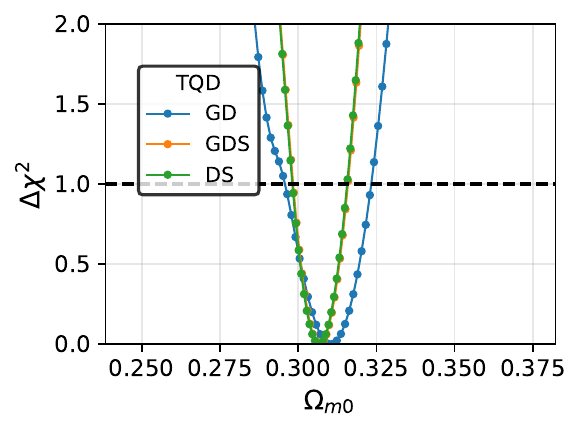}}
    \subfloat[$\delta$\label{3d}]{\includegraphics[width=0.25\textwidth,keepaspectratio]{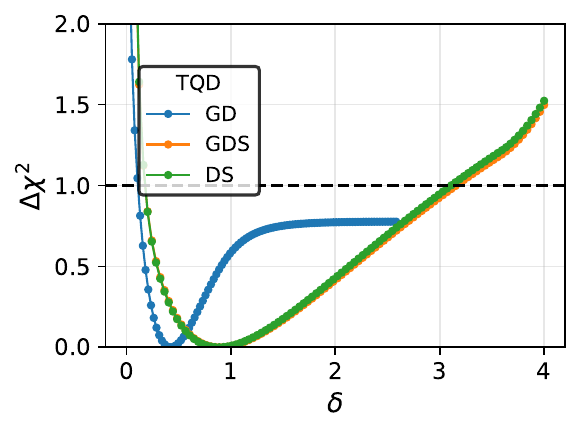}}
    \subfloat[$\gamma$\label{3e}]{\includegraphics[width=0.25\textwidth,keepaspectratio]{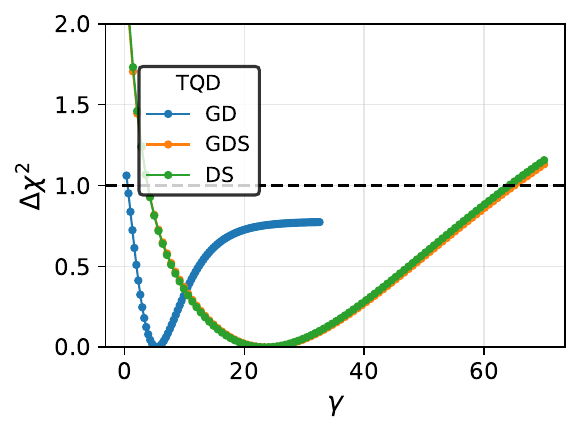}}
    \subfloat[$\sigma_8$\label{3f}]{\includegraphics[width=0.25\textwidth,keepaspectratio]{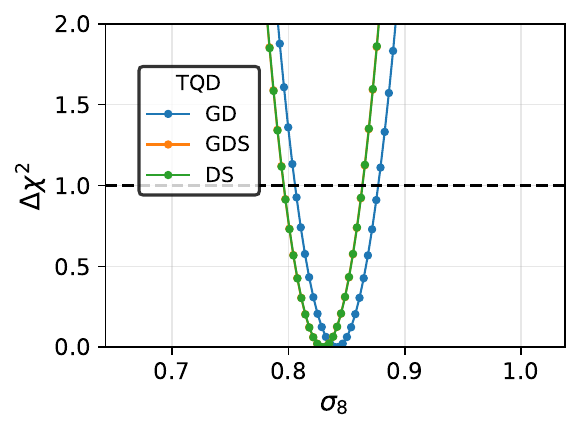}}
    \caption{Bayesian and frequentist constraints on the TQD model. Top row: Bayesian posterior contours (Fig.~\ref{3a}) and the corresponding EOS (Fig.~\ref{3b}); contours represent 68\% and 95\% credible regions. Bottom row: PL constraints on $\Omega_{m0}$ (Fig.~\ref{3c}), $\delta$ (Fig.~\ref{3d}), $\gamma$ (Fig.~\ref{3e}) and $\sigma_8$ (Fig.~\ref{3f}). The $\Delta \chi^2 = 1$ line marks the 68\% confidence interval.}
    \label{fig3}
\end{figure}

\begin{figure}[ht]
    \centering
    \subfloat[Bayesian constraints on GEDE \label{4a}]{\includegraphics[width=0.6\textwidth,keepaspectratio]{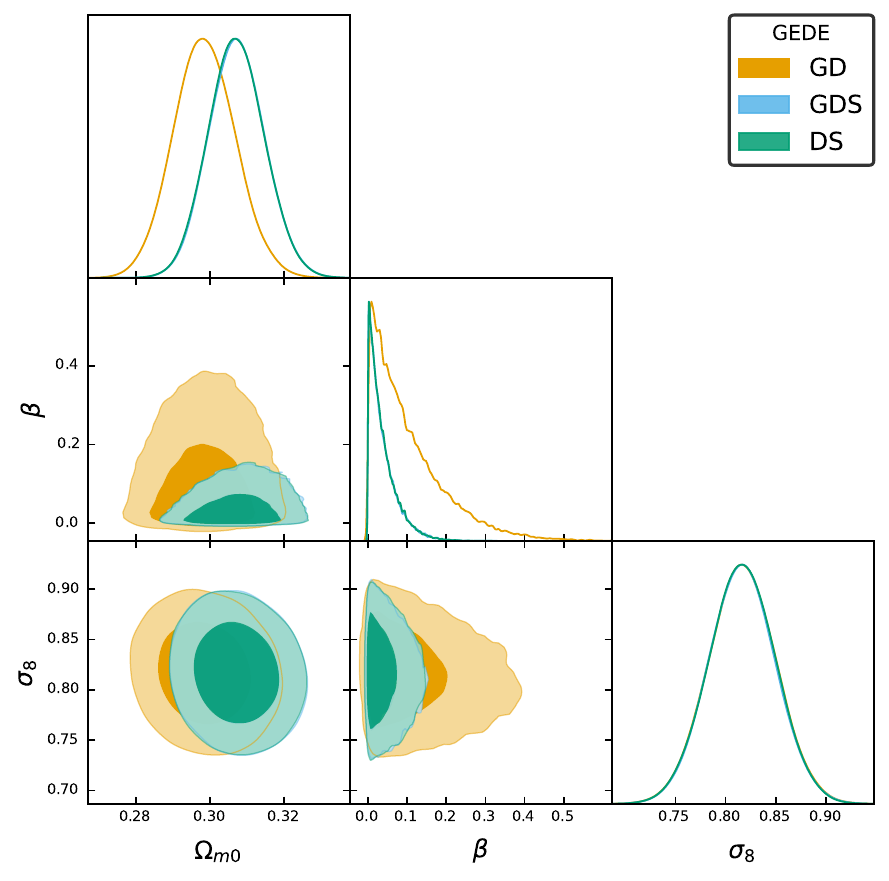}}
    \subfloat[GEDE \label{4b}]{\includegraphics[width=0.4\textwidth,keepaspectratio]{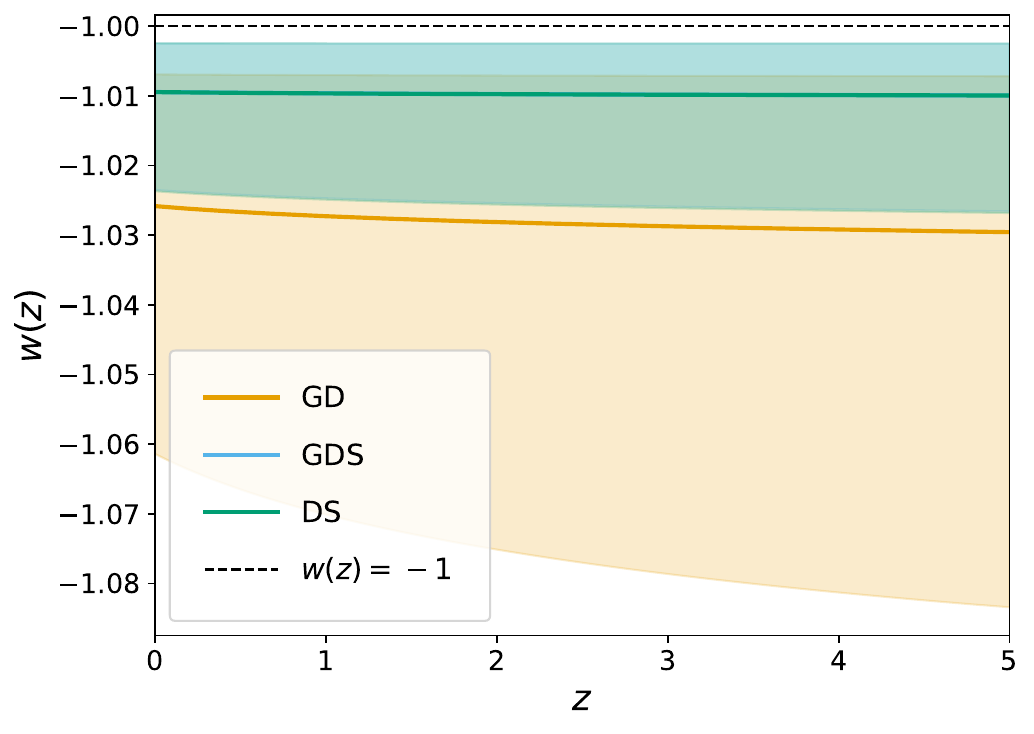}}\\
    \subfloat[$\Omega_{m0}$\label{4c}]{\includegraphics[width=0.33\textwidth,keepaspectratio]{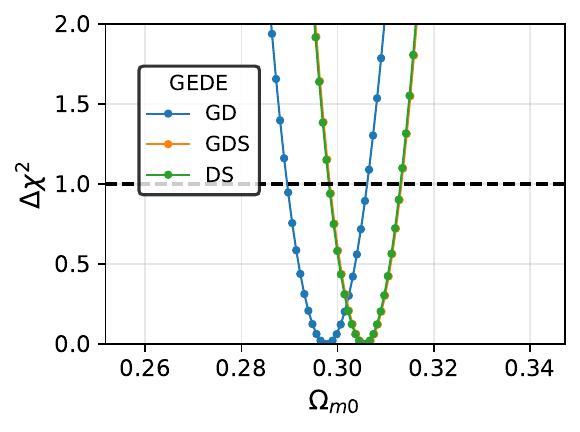}}
    \subfloat[$\beta$\label{4d}]{\includegraphics[width=0.33\textwidth,keepaspectratio]{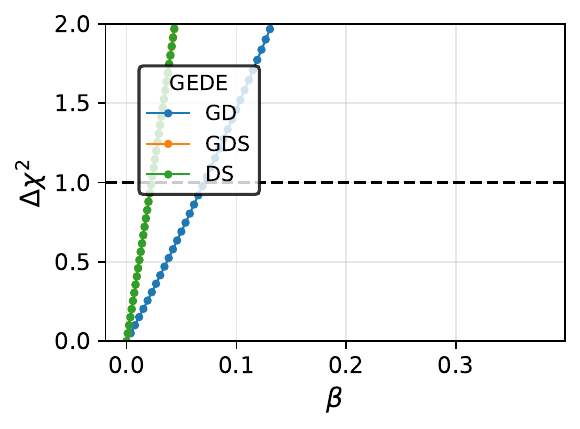}}
    \subfloat[$\sigma_8$\label{4e}]{\includegraphics[width=0.33\textwidth,keepaspectratio]{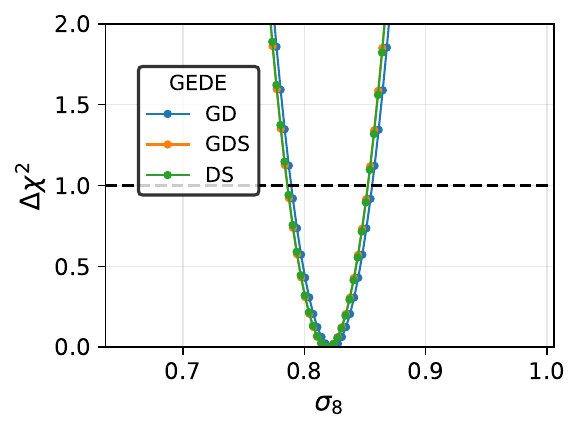}}
    \caption{Bayesian and frequentist constraints on the GEDE model. Top row: Bayesian posterior contours (Fig.~\ref{4a}) and the corresponding EOS (Fig.~\ref{4b}); contours represent 68\% and 95\% credible regions. Bottom row: PL constraints on $\Omega_{m0}$ (Fig.~\ref{4c}), $\beta$ (Fig.~\ref{4d}) and $\sigma_8$ (Fig.~\ref{4e}). The $\Delta \chi^2 = 1$ line marks the 68\% confidence interval; for $\beta$, which is bounded below by 0, the 95\% upper limit is determined using $\Delta \chi^2 = 2.71$, as discussed in the text.}
    \label{fig4}
\end{figure}

\begin{figure}[ht]
    \centering
    \subfloat[Bayesian constraints on AT \label{5a}]{\includegraphics[width=0.6\textwidth,keepaspectratio]{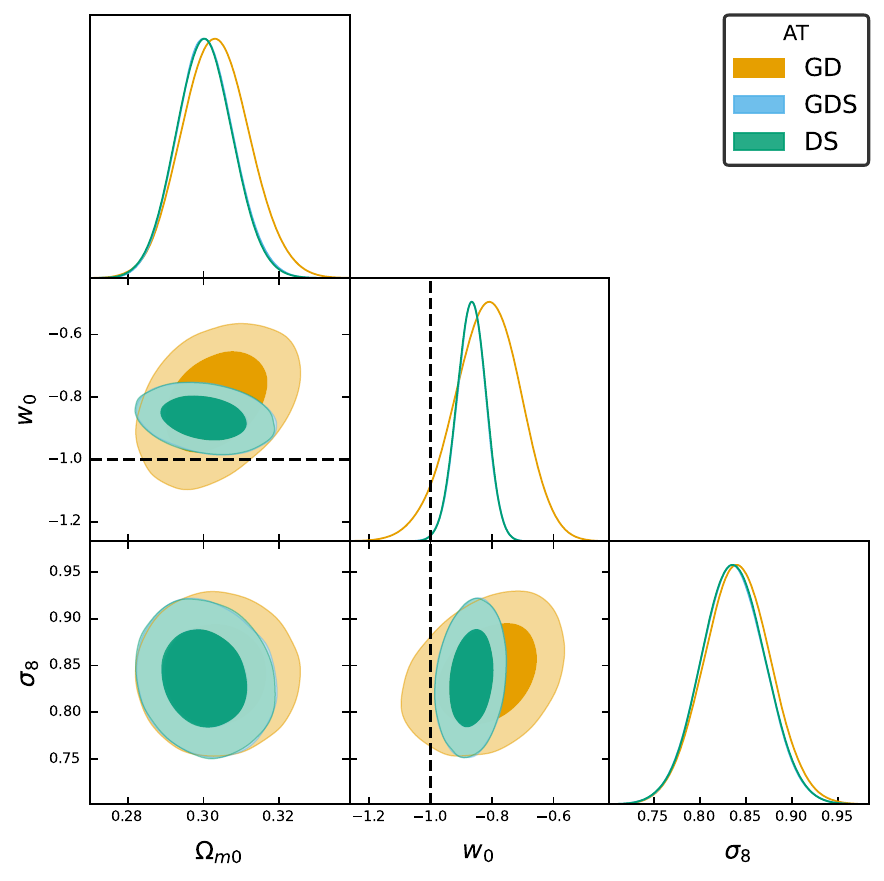}}
    \subfloat[AT \label{5b}]{\includegraphics[width=0.4\textwidth,keepaspectratio]{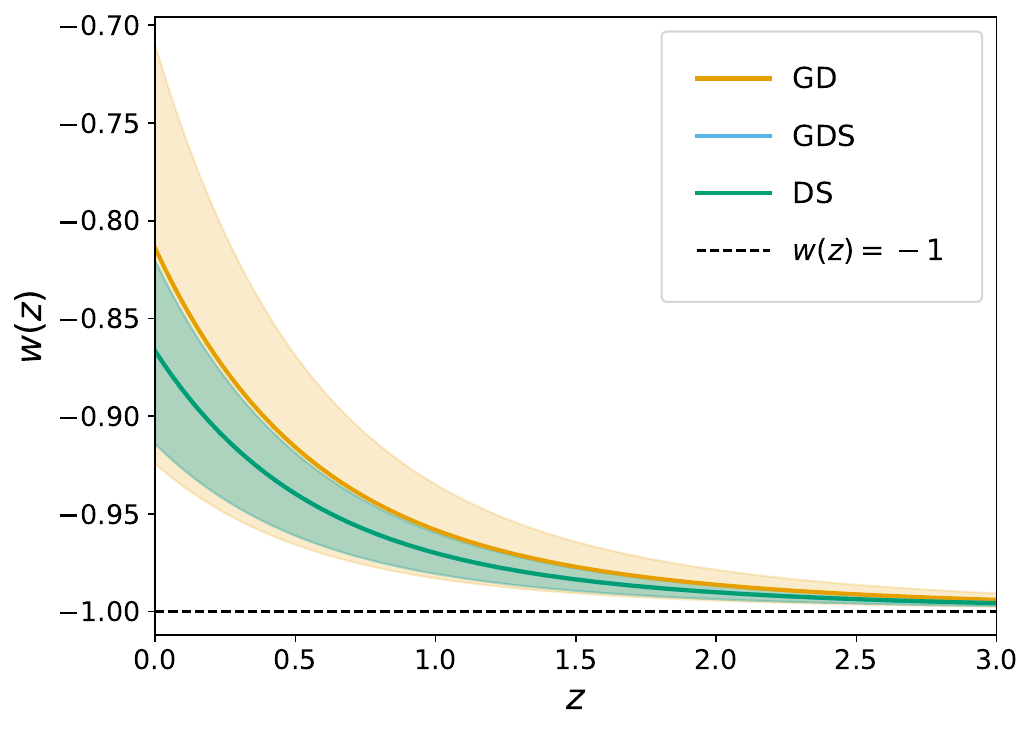}}\\
    \subfloat[$\Omega_{m0}$\label{5c}]{\includegraphics[width=0.33\textwidth,keepaspectratio]{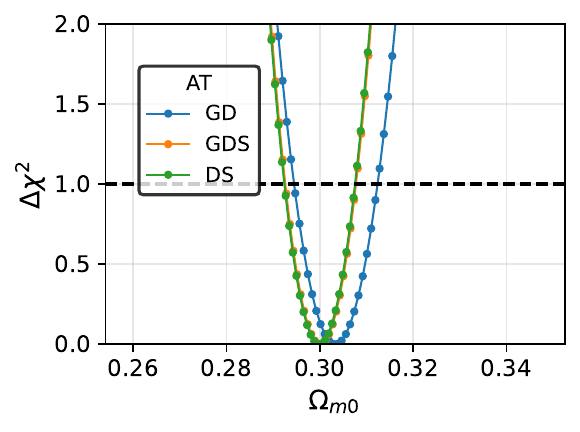}}
    \subfloat[$w_0$\label{5d}]{\includegraphics[width=0.33\textwidth,keepaspectratio]{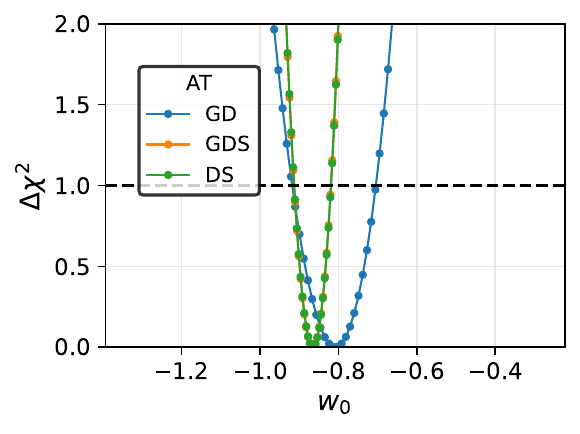}}
    \subfloat[$\sigma_8$\label{5e}]{\includegraphics[width=0.33\textwidth,keepaspectratio]{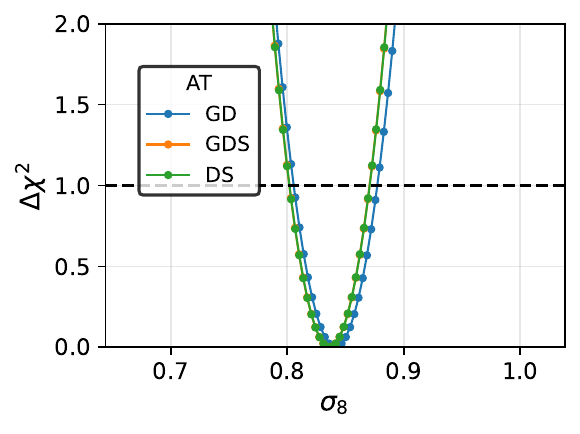}}
    \caption{Bayesian and frequentist constraints on the AT model. Top row: Bayesian posterior contours  (Fig.~\ref{5a}) and the corresponding EOS (Fig.~\ref{5b}); contours represent 68\% and 95\% credible regions. Bottom row: PL constraints on $\Omega_{m0}$ (Fig.~\ref{5c}), $w_0$ (Fig.~\ref{5d}) and $\sigma_8$ (Fig.~\ref{5e}). The $\Delta \chi^2 = 1$ line marks the 68\% confidence interval.}
    \label{fig5}
\end{figure}

\begin{figure}[ht]
    \centering
    \subfloat[$\ln B_{1i}$ \label{6a}]{\includegraphics[width=0.33\textwidth,keepaspectratio]{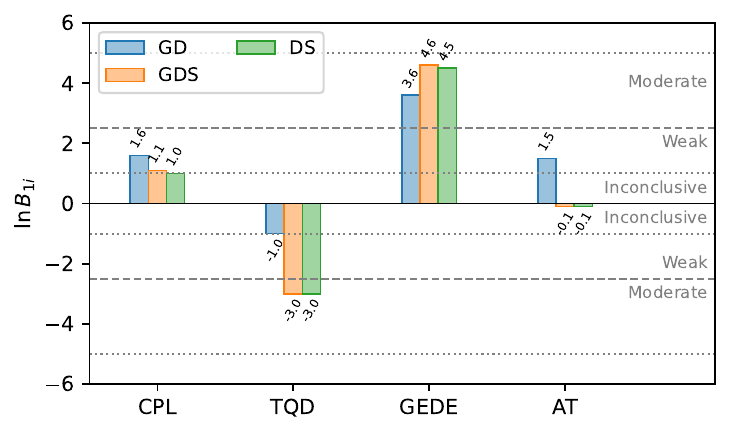}}
    \subfloat[$\Delta$AIC \label{6b}]{\includegraphics[width=0.33\textwidth,keepaspectratio]{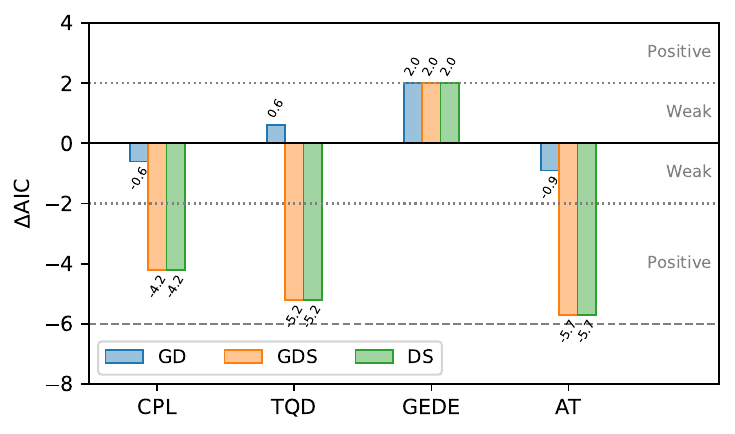}}
    \subfloat[$\Delta$DIC \label{6c}]{\includegraphics[width=0.33\textwidth,keepaspectratio]{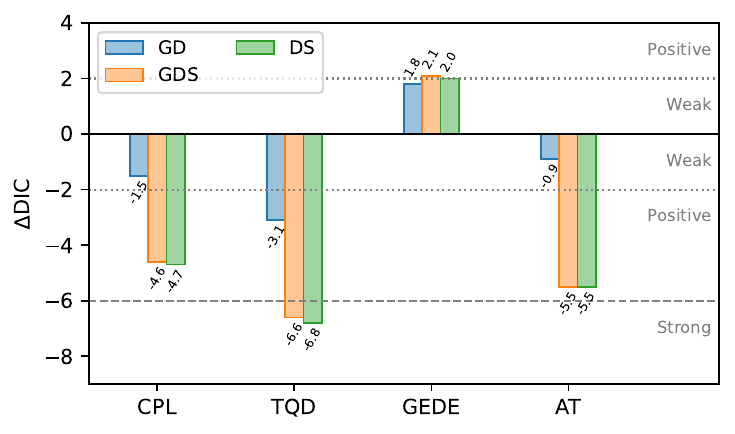}}
    \caption{Model comparison results relative to $\Lambda$CDM (the reference model). Positive values of $\ln B_{1i}$, $\Delta$AIC and $
    \Delta$DIC favor $\Lambda$CDM, whereas negative values favor the alternative DE parameterization. Horizontal lines and labels indicate the corresponding strength-of-evidence thresholds (cf. Section~\ref{sec3}).}
    \label{fig6}
\end{figure}

The 68\% Bayesian credible intervals and frequentist confidence intervals for the cosmological parameters are given in Tables~\ref{table3}--\ref{table7}. Figs.~\ref{fig1}--\ref{fig5} show the corresponding posterior distributions, the frequentist PLs for the different EOS. Model comparison based on Bayesian evidence, AIC and DIC are shown in Fig.~\ref{fig6}. We also the EOS (and the 68\% credible bands) of the different model frameworks, reconstructed from the corresponding MCMC samples.

First, we consider the parameter constraints on $\Lambda$CDM (Table~\ref{table3}, Fig.~\ref{fig1}). We find that the Bayesian and frequentist constraints agree well within $1\sigma$. Including SNe data into the dataset combination shifts the constraint on $\Omega_{m0}$ from $0.2980\pm0.0083$ in GD to $0.3059\pm0.0075$ in GDS, with a $\sim 10\%$ reduction in the uncertainty. In Ref.~\cite{desi_2025}, $\Omega_{m0}$ was reported as $0.2975\pm0.0086$ (BAO only) which is in close agreement with our GD result. This suggests that the BAO data dominate the constraints, while the GRB data contribute only marginally. We obtain $\sigma_8 = 0.819\pm0.033$ (GDS), which is in close agreement with the CMB inferred value (ACT DR6+Planck gives $\sigma_8 = 0.8149\pm0.0063$ \cite{act_2025}, a shift of $0.1\sigma$). A recent review of the status of the $S_8$ tension can be found in Ref.~\cite{Pantos_2026}. The values of $\sigma_8$ obtained from different dataset combinations are consistent within $1\sigma$. This is expected, as GRBs, BAO and SNe primarily probe the geometry of the Universe and therefore do not significantly alter the constraints on structure growth provided by growth-related data. 

\begin{table}[ht]
\caption{68\% Bayesian credible intervals and frequentist confidence intervals for $\Lambda$CDM.}
\label{table3}
\centering

\setlength{\tabcolsep}{3pt}

\begin{tabular}{c|cc|cc|cc}
\hline
\hline

\multirow{2}{*}{\thead{Parameters}} & \multicolumn{2}{c|}{\thead{GD}} & \multicolumn{2}{c|}{\thead{DS}} & \multicolumn{2}{c}{\thead{GDS}} \\
 & \thead{Bayesian} & \thead{Frequentist} & \thead{Bayesian} & \thead{Frequentist} & \thead{Bayesian} & \thead{Frequentist} \\

\hline
\hline
&  &  &  &  &  &\\[1ex]
$H_0r_d$ [km/s] & $10160\pm69$ & $10161^{+66}_{-73}$ & $10100\pm62$ & $10102^{+59}_{-64}$ & $10100\pm62$ & $10100^{+59}_{-64}$ \\[0.7ex]
$\Omega_{m0}$ & $0.2980\pm0.0083$ & $0.2973^{+0.0089}_{-0.0078}$ & $0.3057\pm0.0074$ & $0.3052^{+0.0079}_{-0.0070}$ & $0.3059\pm0.0075$ & $0.3054^{+0.0079}_{-0.0070}$ \\[0.7ex]
$\sigma_8$  & $0.823\pm0.034$ & $0.821^{+0.035}_{-0.032}$ & $0.819\pm0.033$ & $0.821^{+0.032}_{-0.035}$ & $0.819\pm0.033$ & $0.818^{+0.035}_{-0.032}$ \\[0.7ex]
\hline
&  &  &  &  &  &\\[0.5ex]
$\Delta \chi^2(\bar{\theta})$ & $0$ & $-$ & $0$ & $-$ & $0$ & $-$ \\[0.7ex]
$\Delta p_D$ & $0$ & $-$ & $0$ & $-$ & $0$ & $-$ \\[0.7ex]
$\chi^2_\mathrm{min}$ & \multicolumn{2}{c|}{$97.2$} & \multicolumn{2}{c|}{$1621.8$} & \multicolumn{2}{c}{$1741.1$} \\[0.7ex]
$\Delta \chi^2_\mathrm{min}$ & \multicolumn{2}{c|}{$0$} & \multicolumn{2}{c|}{$0$} & \multicolumn{2}{c}{$0$} \\[0.7ex]
\hline
\hline

\end{tabular}
\end{table}

Next, we consider the CPL DE parameterization. The results are presented in Table~\ref{table4} and Fig.~\ref{fig2}. The Bayesian and frequentist constraints agree within $1\sigma$. We also do not observe any significant shift in $\Omega_{m0}$ across all dataset combinations. Ref.~\cite{desi_2025} reports $\Omega_{m0} = 0.352^{+0.041}_{-0.018}$ (BAO only) while our GD result is $\Omega_{m0} = 0.3260^{+0.0330}_{-0.0130}$, corresponding to a $0.70\sigma$ downward shift. Ref.~\cite{desi_2025} also reports $(w_0, w_a) = (-0.48^{+0.35}_{-0.17}, <-1.34)$ which agrees with our GD results $(-0.57^{+0.30}_{-0.21}, -1.34^{+0.63}_{-1.30})$. This again suggests that the GRB data contribute only marginally to the parameter constraints. Including SNe moves the central $w_0$ and $w_a$ values closer to -1 and 0, respectively ($w_0 = -0.83^{+0.07}_{-0.08}$ and $w_a = -0.47^{+0.47}_{-0.58}$ in GDS). Because the errors on these parameters are large, the shift between GD and GDS is small ($\leq 1\sigma$). The errors decrease by at least $20\%$ when SNe are included, indicating that the SNe data tighten the constraints on the background parameters. The CPL value of $\sigma_8=0.834^{+0.032}_{-0.046}$ in GDS is consistent with the $\Lambda$CDM values $0.819\pm0.033$. The $w_0$ values differ from -1 by $\sim 2\sigma$, while $w_a$ is consistent with 0 at $1\sigma$. This can be seen in the contour plots in Fig.~\ref{fig2}. The EOS (Fig.~\ref{2b}) shows phantom crossing at $z\sim0.5$. The crossing occurs at a significance of $\sim (2-3) \sigma$. We also find relatively weak constraints on $w_a$ for DS and GDS, with characteristic uncertainties of order $\sim 0.5$, compared with $\sim 0.2$ in analyses incorporating CMB data alongside BAO and SNe~\cite{desi_2025}. Our data is sensitive to $w_a$ only up to $z\sim2$, while CMB adds a high-redshift anchor. 

\begin{table}[ht]
\caption{68\% Bayesian credible intervals and frequentist confidence intervals for CPL.}
\label{table4}
\centering

\setlength{\tabcolsep}{3pt}

\begin{tabular}{c|cc|cc|cc}
\hline
\hline

\multirow{2}{*}{\thead{Parameters}} & \multicolumn{2}{c|}{\thead{GD}} & \multicolumn{2}{c|}{\thead{DS}} & \multicolumn{2}{c}{\thead{GDS}} \\
 & \thead{Bayesian} & \thead{Frequentist} & \thead{Bayesian} & \thead{Frequentist} & \thead{Bayesian} & \thead{Frequentist} \\
 
\hline
\hline
&  &  &  &  &  & \\[1ex]
$H_0r_d$ [km/s] & $9474^{+190}_{-240}$ & $9664^{+240}_{-206}$ & $9995\pm72$ & $9993^{+76}_{-68}$ & $9994\pm72$ & $9999^{+69}_{-75}$ \\[0.7ex]
$\Omega_{m0}$ & $0.3260^{+0.0330}_{-0.0130}$ & $0.3396^{+0.0206}_{-0.0261}$ & $0.3040^{+0.0210}_{-0.0080}$ & $0.3096^{+0.0125}_{-0.0143}$ & $0.3030^{+0.0220}_{-0.0081}$ & $0.3106^{+0.0112}_{-0.0158}$ \\[0.7ex]
$w_0$ & $-0.57^{+0.30}_{-0.21}$ & $-0.47^{+0.26}_{-0.29}$ & $-0.83^{+0.07}_{-0.08}$ & $-0.83^{+0.08}_{-0.07}$ & $-0.83^{+0.07}_{-0.08}$ & $-0.82^{+0.07}_{-0.08}$ \\[0.7ex] 
$w_a$ & $-1.34^{+0.63}_{-1.30}$ & $-1.70^{+1.03}_{-1.02}$ & $-0.50^{+0.46}_{-0.57}$ & $-0.59^{+0.49}_{-0.46}$ & $-0.47^{+0.47}_{-0.58}$ & $-0.55^{+0.47}_{-0.49}$ \\[0.7ex] 
$\sigma_8$ & $0.835^{+0.033}_{-0.042}$ & $0.829^{+0.033}_{-0.036}$ & $0.833^{+0.032}_{-0.045}$ & $0.826^{+0.035}_{-0.037}$ & $0.834^{+0.032}_{-0.046}$ & $0.823^{+0.038}_{-0.034}$ \\[0.7ex]
\hline
&  &  &  &  &  &\\[0.5ex]
$\Delta \chi^2(\bar{\theta})$ & $-4.1$ & $-$ & $-7.3$ & $-$ & $-7.2$ & $-$ \\[0.7ex]
$\Delta p_D$ & $1.3$ & $-$ & $1.3$ & $-$ & $1.3$ & $-$ \\[0.7ex]
$\chi^2_\mathrm{min}$ & \multicolumn{2}{c|}{$92.6$} & \multicolumn{2}{c|}{$1613.6$} & \multicolumn{2}{c}{$1732.9$} \\[0.7ex]
$\Delta \chi^2_\mathrm{min}$ & \multicolumn{2}{c|}{$-4.6$} & \multicolumn{2}{c|}{$-8.2$} & \multicolumn{2}{c}{$-8.2$} \\ [0.7ex]
\hline
\hline

\end{tabular}
\end{table}

For the TQD parameterization (Table~\ref{table5}, Fig.~\ref{fig3}) , the $\Lambda$CDM model is recovered in the limit $\delta \rightarrow0$. From Table~\ref{table5}, we find a mild deviation in Bayesian constraints from $\delta = 0$ ($2.2 \sigma$ and $2.1\sigma$ for DS and GDS, respectively). The frequentist constraints on $\delta$ (for DS and GDS) are within $1.3\sigma$ of 0. The Bayesian posterior of $\gamma$ increases monotonically (Fig.~\ref{3a}) for DS and GDS, so we quote $68\%$ and $95\%$ lower limits. In contrast, frequentist analysis gives constraints on $\gamma$ with wide error bars. For GD, the Bayesian analysis gives weak constraints on $\delta$ which can be seen from the $\sim 40\%$ reduction in the error bars when SNe dataset is included (moving from GD to DS or GDS). We do not quote constraints on $\delta$ and $\gamma$ from the frequentist analysis (for the GD dataset combination) because the PL crosses $\Delta\chi^2=1$ only on the lower side. On the upper side, it runs parallel to $\Delta\chi^2=1$. On neither side does it reach $\Delta\chi^2= 2.71$ within the range considered (Figs.~\ref{3d} and~\ref{3e}). Hence, neither a $68\%$ interval nor a $95\%$ limit can be defined. The frequentist and Bayesian constraints on $\delta$ are consistent within $1\sigma$ but the $1\sigma$ error bars in the frequentist analysis are quite large. There is a wedge-shaped degeneracy between the positively correlated $\delta$ and $\gamma$. Using CMB$+$BAO$+$SNe, Ref.~\cite{souza_2025} found $\delta = 0.131^{+0.054}_{-0.10}$ and $\gamma > 5.420$ at $1\sigma$. Our Bayesian analysis shows a $\sim 1.3 \sigma$ tension in $\delta$ with this result. Including CMB data also reduces the error on $\delta$ by $\sim 56\%$.

The EOS curve for the TQD parameterization (Fig.~\ref{3b}) shows a rapid decline to reach $w(z)=-1$ asymptotically. Fig. 2 of Ref.~\cite{souza_2025} also shows a rapid decline in the EOS curve. Their results showed $w(z \gtrsim 0.4) \sim -1$ while we get a similar $w(z\gtrsim0.3) \sim -1$ (for DS and GDS) and $w(z \gtrsim 0.5) \sim -1$ for GD. 

\begin{table}[ht]
\caption{68\% Bayesian credible intervals and frequentist confidence intervals for TQD. We quote lower limits on $\gamma$ as ($68\%, 95\%$) for the DS and GDS cases in the Bayesian constraints.}
\label{table5}
\centering

\setlength{\tabcolsep}{3pt}

\begin{tabular}{c|cc|cc|cc}
\hline
\hline

\multirow{2}{*}{\thead{Parameters}} & \multicolumn{2}{c|}{\thead{GD}} & \multicolumn{2}{c|}{\thead{DS}} & \multicolumn{2}{c}{\thead{GDS}} \\
 & \thead{Bayesian} & \thead{Frequentist} & \thead{Bayesian} & \thead{Frequentist} & \thead{Bayesian} & \thead{Frequentist} \\

\hline
\hline
&  &  &  &  &  & \\[1ex]
$H_0r_d$ [km/s] & $9868\pm150$ & $9827^{+194}_{-209}$ & $9985\pm71$ & $9966^{+80}_{-108}$ & $9984\pm71$ & $9956^{+89}_{-100}$ \\[0.7ex]
$\Omega_{m0}$ & $0.3090\pm0.0100$ & $0.3097^{+0.0136}_{-0.0139}$ & $0.3040\pm0.0076$ & $0.3062^{+0.0094}_{-0.0081}$ & $0.3042\pm0.0076$ & $0.3073^{+0.0086}_{-0.0089}$ \\[0.7ex]
$\delta$ & $0.51^{+0.27}_{-0.32}$ & $-$ & $0.37^{+0.14}_{-0.20}$ & $0.89^{+2.22}_{-0.71}$ & $0.37^{+0.14}_{-0.21}$ & $0.89^{+2.27}_{-0.71}$ \\[0.7ex] 
$\gamma$ & $8.6^{+5.7}_{-3.3}$ & $-$ & $(> 11.52, >2.88)$ & $23.33^{+40.82}_{-19.45}$ & $(>11.52, >2.83)$ & $24.04^{+40.98}_{-20.16}$ \\[0.7ex]
$\sigma_8$ & $0.837\pm0.035$ & $0.843^{+0.034}_{-0.037}$ & $0.832\pm0.034$ & $0.828^{+0.035}_{-0.032}$ & $0.831\pm0.034$ & $0.828^{+0.035}_{-0.032}$\\[0.7ex]
\hline
&  &  &  &  &  &\\[0.5ex]
$\Delta \chi^2(\bar{\theta})$ & $-1.9$ & $-$ & $-8.8$ & $-$ & $-8.8$ & $-$ \\[0.7ex]
$\Delta p_D$ & $-0.6$ & $-$ & $1.0$ & $-$ & $1.1$ & $-$ \\[0.7ex]
$\chi^2_\mathrm{min}$ & \multicolumn{2}{c|}{$93.8$} & \multicolumn{2}{c|}{$1612.6$} & \multicolumn{2}{c}{$1731.9$} \\[0.7ex]
$\Delta \chi^2_\mathrm{min}$ & \multicolumn{2}{c|}{$-3.4$} & \multicolumn{2}{c|}{$-9.2$} & \multicolumn{2}{c}{$-9.2$} \\ [0.7ex]
\hline
\hline

\end{tabular}
\end{table}

For the GEDE parameterization (Table~\ref{table6}, Fig.~\ref{fig4}), deviations from $\beta = 0$ correspond to deviations from $\Lambda$CDM. Ref.~\cite{yang_2021} found that Planck 2018 data alone exclude $\Lambda$CDM at more than $95\%$ confidence level and the Phenomenologically emergent DE (PEDE) model at more than $68\%$ confidence level. Including BAO data restores consistency with $\Lambda$CDM within $1\sigma$ and with PEDE within $2\sigma$. In our analysis, we adopted the prior $\beta \in [0, 4.34]$, following~\cite{yang_2026}, to suppress the high-redshift DE density relative to $\Lambda$CDM. We obtain $95\%$ upper limits of $\beta < 0.3, 0.11$ and 0.11 for the GD, DS and GDS datasets, respectively. These limits include $\beta=0$ so we find no exclusion of $\Lambda$CDM. The PEDE model corresponds to $\beta=0.434$, which lies above the $95\%$ upper limits for all three dataset combinations. Hence, PEDE is excluded at more than $95\%$ in our analysis. We find that including SNe data tightens the upper limit on $\beta$. Furthermore, the frequentist $95\%$ upper limits are lower than the corresponding Bayesian upper limits. $\Omega_{m0}$ remains consistent within $1\sigma$ across all dataset combinations considered. The EOS (Fig.~\ref{4b}) shows $w(z)<-1$ at all redshifts (see also Fig. 2 of Ref.~\cite{li_2020}). This parameterization does not show any crossing of $w(z) = -1$. This behavior can be understood from Eqn.~\ref{eqn6}. For $\beta\geq0$, $-1-\frac{2\beta}{3} \leq w(z) \leq -1$. To investigate the constraints on deviations from $\Lambda$CDM without restricting $\beta$ to non-negative values, we relax the prior in Appendix~\ref{appA} and repeat the analysis.

\begin{table}[ht]
\caption{68\% Bayesian credible intervals and frequentist confidence intervals for GEDE.}
\label{table6}
\centering

\setlength{\tabcolsep}{3pt}

\begin{tabular}{c|cc|cc|cc}
\hline
\hline

\multirow{2}{*}{\thead{Parameters}} & \multicolumn{2}{c|}{\thead{GD}} & \multicolumn{2}{c|}{\thead{DS}} & \multicolumn{2}{c}{\thead{GDS}} \\
 & \thead{Bayesian} & \thead{Frequentist} & \thead{Bayesian} & \thead{Frequentist} & \thead{Bayesian} & \thead{Frequentist} \\

\hline
\hline
&  &  &  &  &  & \\[1ex]
$H_0r_d$ [km/s] & $10210^{+76}_{-91}$ & $10161^{+67}_{-73}$ & $10110\pm63$ & $10102^{+59}_{-65}$ & $10110\pm63$ & $10094^{+65}_{-58}$ \\[0.7ex]
$\Omega_{m0}$ & $0.2986\pm0.0083$ & $0.2981^{+0.0080}_{-0.0086}$ & $0.3072\pm0.0076$ & $0.3052^{+0.0079}_{-0.0070}$ & $0.3073\pm0.0076$ & $0.3061^{+0.0071}_{-0.0077}$ \\[0.7ex]
$\beta$ & $<0.30$ & $<0.17$ & $<0.11$ & $<0.06$ & $<0.11$ & $<0.06$ \\[0.7ex] 
$\sigma_8$ & $0.817\pm0.034$ & $0.821^{+0.035}_{-0.032}$ & $0.817\pm0.033$ & $0.821^{+0.032}_{-0.034}$ & $0.816\pm0.033$ & $0.821^{+0.032}_{-0.035}$ \\[0.7ex]
\hline
&  &  &  &  &  &\\[0.5ex]
$\Delta \chi^2(\bar{\theta})$ & $1.6$ & $-$ & $1.8$ & $-$ & $1.7$ & $-$ \\[0.7ex]
$\Delta p_D$ & $0.1$ & $-$ & $0.1$ & $-$ & $0.2$ & $-$ \\[0.7ex]
$\chi^2_\mathrm{min}$ & \multicolumn{2}{c|}{$97.2$} & \multicolumn{2}{c|}{$1621.8$} & \multicolumn{2}{c}{$1741.1$} \\[0.7ex]
$\Delta \chi^2_\mathrm{min}$ & \multicolumn{2}{c|}{$0$} & \multicolumn{2}{c|}{$0$} & \multicolumn{2}{c}{$0$} \\ [0.7ex]
\hline
\hline

\end{tabular}
\end{table}

Ref.~\cite{lodha_2024} considered the AT parameterization with $b=0.5$ and $p=1$ (similar to us) and confronted it to DESI DR1 BAO$+$SNe$+$CMB dataset combination. They reported $w_0 = -1.17\pm0.15$ for CMB$+$BAO and $w_0 = -0.954\pm0.040$ for CMB$+$BAO$+$SNe (PantheonPlus compilation). On the other hand, Ref.~\cite{lodha_2025} allowed $w_0$ and $p$ to both be free parameters and found that the posteriors for $p$ are largely unconstrained and hit the prior boundary $p=30$. In our analysis (Table~\ref{table7}, Fig.~\ref{fig5}), we find that even though the $w_0$ values across all three dataset combinations are consistent within $1\sigma$, the error bars show a $\sim 50\%$ reduction upon the  inclusion of the SNe dataset. We also notice a $\sim1.7\sigma$ (GD) and $\sim2.6\sigma$ (DS and GDS) discrepancy with $w_0 = -1$ (Fig.~\ref{fig5}). Furthermore, comparing our results with Ref.~\cite{lodha_2024}, we find a $1.9\sigma$ tension in $w_0$ values (GDS vs. CMB$+$BAO) and a $1.31\sigma$ tension (GDS vs. CMB$+$BAO$+$SNe). The EOS curve is shown in Fig.~\ref{5b} where, compared to TQD, the curve reaches the asymptotic $w(z) = -1$ gradually. This gradual decrease is because we fixed $p = 1$. We look at a more rapid decline ($p = 20$) in Appendix~\ref{appB}. It is worth noting here that, for $p=1$, AT gives similar constraints to the calibrated thawing form $w_a \approx -1.58(1+w_0)$, while remaining non-phantom at all times~\cite{lodha_2024, putter_2008}.

\begin{table}[ht]
\caption{68\% Bayesian credible intervals and frequentist confidence intervals for AT.}
\label{table7}
\centering

\setlength{\tabcolsep}{3pt}

\begin{tabular}{c|cc|cc|cc}
\hline
\hline

\multirow{2}{*}{\thead{Parameters}} & \multicolumn{2}{c|}{\thead{GD}} & \multicolumn{2}{c|}{\thead{DS}} & \multicolumn{2}{c}{\thead{GDS}} \\
 & \thead{Bayesian} & \thead{Frequentist} & \thead{Bayesian} & \thead{Frequentist} & \thead{Bayesian} & \thead{Frequentist} \\

\hline
\hline
&  &  &  &  &  & \\[1ex]
$H_0r_d$ [km/s] & $9912^{+150}_{-170}$ & $9910^{+153}_{-161}$ & $9991\pm71$ & $9995^{+68}_{-75}$ & $9990\pm71$ & $9994^{+68}_{-74}$ \\[0.7ex]
$\Omega_{m0}$ & $0.3033\pm0.0089$ & $0.3029^{+0.0095}_{-0.0084}$ & $0.3003\pm0.0076$ & $0.2996^{+0.0080}_{-0.0072}$ & $0.3005\pm0.0076$ & $0.2997^{+0.0080}_{-0.0071}$ \\[0.7ex]
$w_0$ & $-0.82^{+0.11}_{-0.10}$ & $-0.81^{+0.11}_{-0.10}$ & $-0.87\pm0.05$ & $-0.86^{+0.04}_{-0.05}$ & $-0.87\pm0.05$ & $-0.87\pm0.05$ \\[0.7ex] 
$\sigma_8$ & $0.841\pm0.036$ & $0.843^{+0.034}_{-0.038}$ & $0.836\pm0.035$ & $0.835^{+0.036}_{-0.033}$ & $0.836\pm0.035$ & $0.834^{+0.036}_{-0.033}$ \\[0.7ex]
\hline
&  &  &  &  &  &\\[0.5ex]
$\Delta \chi^2(\bar{\theta})$ & $-2.9$ & $-$ & $-7.7$ & $-$ & $-7.7$ & $-$ \\[0.7ex]
$\Delta p_D$ & $1.0$ & $-$ & $1.1$ & $-$ & $1.1$ & $-$ \\[0.7ex]
$\chi^2_\mathrm{min}$ & \multicolumn{2}{c|}{$94.3$} & \multicolumn{2}{c|}{$1614.1$} & \multicolumn{2}{c}{$1733.4$} \\[0.7ex]
$\Delta \chi^2_\mathrm{min}$ & \multicolumn{2}{c|}{$-2.9$} & \multicolumn{2}{c|}{$-7.7$} & \multicolumn{2}{c}{$-7.7$} \\ [0.7ex]
\hline
\hline

\end{tabular}
\end{table}

Next, we compare the parameterizations using Bayes factor, AIC and DIC. The strength of preference has already been described in Sec.~\ref{sec3}.

\textbf{Bayes factor}. From Fig.~\ref{6a}, $\Lambda$CDM is preferred over CPL (weak preference) and GEDE (moderate preference). For AT the evidence is inconclusive when SNe are included, while $\Lambda$CDM is weakly preferred for the GD dataset. In contrast, TQD is moderately preferred over $\Lambda$CDM for the DS and GDS dataset combinations but the evidence is weak for GD.

\textbf{AIC}. From Fig.~\ref{6b}, $\Lambda$CDM is weakly preferred over TQD (for GD) and positively preferred over GEDE (for all dataset combinations). All other parameterization-dataset combinations are weakly or positively preferred over $\Lambda$CDM. The $\Delta \chi^2_\mathrm{min}$ values in Tables~\ref{table4}--\ref{table7} show that CPL, TQD and AT fit the data better than $\Lambda$CDM, whereas GEDE yields an identical best-fit $\chi^2_\mathrm{min}$. Since AIC adds two per extra parameter, a model needs $\Delta \chi^2_\mathrm{min} < -4$ (CPL and TQD) or $\Delta \chi^2_\mathrm{min} < -2$ (AT and GEDE) to be preferred. CPL and AT satisfy this for all dataset combinations, and TQD for DS and GDS. GEDE cannot satisfy it, since its $\Delta \chi^2_\mathrm{min} = 0$, so the AIC penalty always favors $\Lambda$CDM.

\textbf{DIC}. From Fig.~\ref{6c}, CPL and AT are positively favored, while TQD is strongly favored over $\Lambda$CDM for the DS and GDS dataset combinations. For the GD dataset, CPL and AT are weakly favored and TQD is positively favored. $\Lambda$CDM is weakly (GD) and positively (DS and GDS) preferred over GEDE for all dataset combinations. 

Overall, the three metrics mostly agree. All three prefer TQD for DS and GDS and all three prefer $\Lambda$CDM over GEDE. The clear disagreement is for CPL: the Bayes factor weakly prefers $\Lambda$CDM while AIC and DIC prefer CPL (weakly for GD and positively for DS and GDS). For AT, the Bayesian evidence is inconclusive for DS and GDS, while AIC and DIC positively prefer AT. AIC does not depend on the priors, and DIC depends on them only through the posterior, while Bayes factor depends on the full prior volume. This suggests that the differences for CPL and AT may come from the Occam penalty in the Bayes factor~\cite{ong_2025, zapata_2026}. For CPL, the improvement in the best-fit $\chi^2_\mathrm{min}$ ($\Delta \chi^2_\mathrm{min} = -4.6, -8.2$ and $-8.2$ for GD, DS and GDS) corresponds to a gain of $2.3-4.1$ in $\ln B_{1i}$, but the Bayes factor still favors $\Lambda$CDM. This implies an approximate Occam penalty of $\sim(4-5)$ defined as $\ln B_{1i} - \frac{\Delta \chi^2_\mathrm{min}}{2}$ \footnote{These estimates assume $\ln Z \sim \ln \mathrm{(Occam\ factor)} - \chi^2_\mathrm{min}/2$, so that $\ln B_{1i} \sim \mathrm{(Occam\ penalty)} + \Delta \chi^2_\mathrm{min}/2$.}. For AT, the penalty is $\sim(3-4)$ ($\Delta \chi^2_\mathrm{min} = -2.9, -7.7$ and $-7.7$ for GD, DS and GDS), which nearly cancels the fit gain for DS and GDS, leaving the evidence inconclusive. For TQD, the penalty is only $\sim (1-2)$ for DS and GDS, so the fit gain outweighs it and the Bayes factor favors TQD. For GEDE, the best-fit $\chi^2_\mathrm{min}$ is identical to $\Lambda$CDM, so the penalty of $\sim(4-5)$ is not offset by any fit gain and $\Lambda$CDM is preferred. 

Ref.~\cite{souza_2025} found that the TQD parameterization remained competitive with $\Lambda$CDM. In our analysis, however, we find moderate Bayesian preference for TQD over $\Lambda$CDM when SNe data are included (DS and GDS). This difference highlights the sensitivity of Bayesian evidence to the choice of observational datasets and their constraining power. Considering the GEDE parameterization, we got a moderate evidence in favor of $\Lambda$CDM, unlike Ref.~\cite{yang_2021}, where they found that Planck 2018 data drives the preference for GEDE over $\Lambda$CDM. We cannot pinpoint the reason for this difference since we use a prior on $\beta$ which is non-negative and we do not use CMB data. However, we do remove the prior restraint in Appendix~\ref{appA} (allowing negative values of $\beta$) and find inconclusive evidence for either the GEDE or the $\Lambda$CDM model. Ref.~\cite{lodha_2025} showed, using CMB$+$BAO(DR2)$+$SNe, that the DIC metric favors the AT parameterization over $\Lambda$CDM. This is similar to our case (even though we fix $p = 1.0$, unlike their case where $p$ is a free parameter).

\begin{figure}[ht]
    \centering
    \subfloat[CPL\label{7a}]{\includegraphics[width=0.25\textwidth,keepaspectratio]{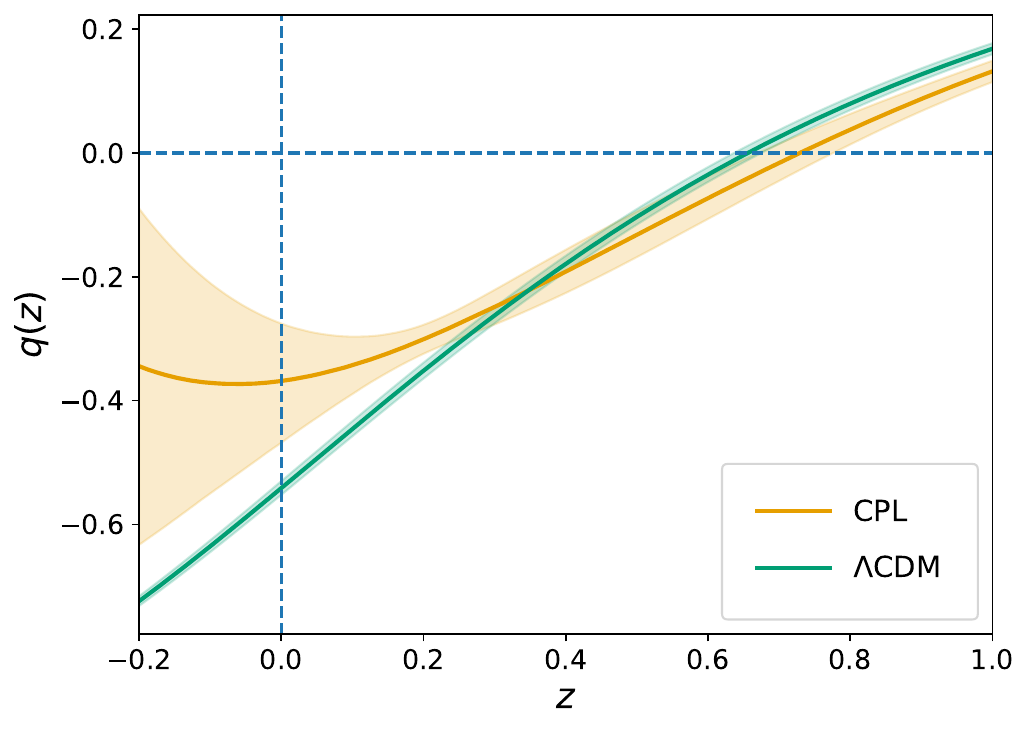}}
    \subfloat[TQD\label{7b}]{\includegraphics[width=0.25\textwidth,keepaspectratio]{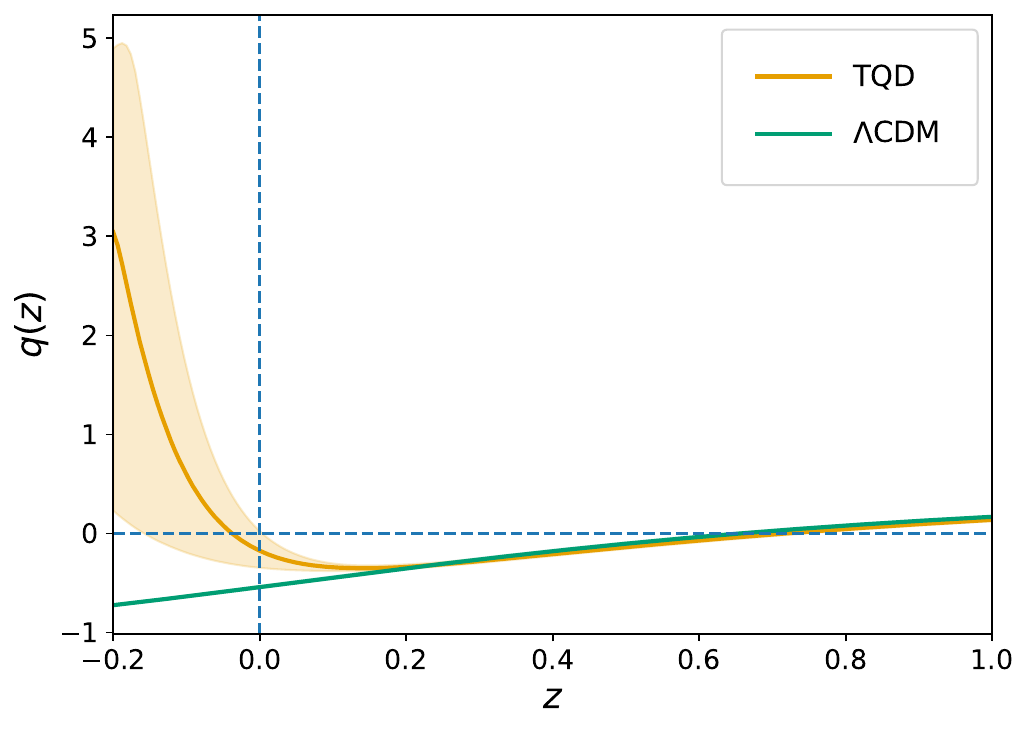}}
    \subfloat[GEDE\label{7c}]{\includegraphics[width=0.25\textwidth,keepaspectratio]{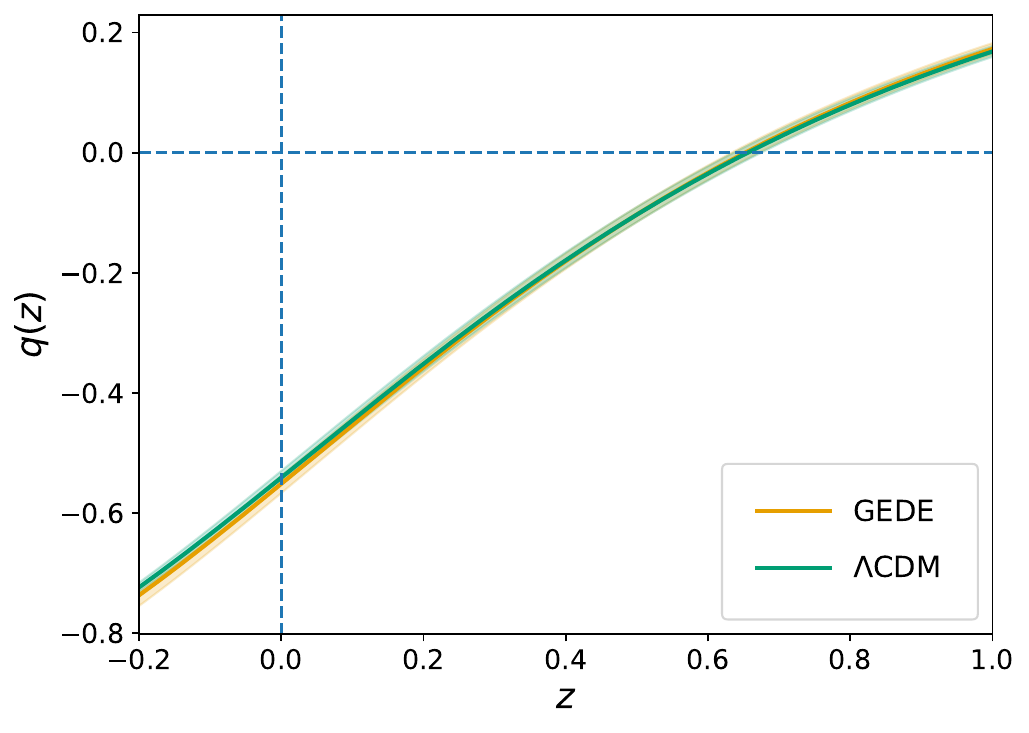}}
    \subfloat[AT\label{7d}]{\includegraphics[width=0.25\textwidth,keepaspectratio]{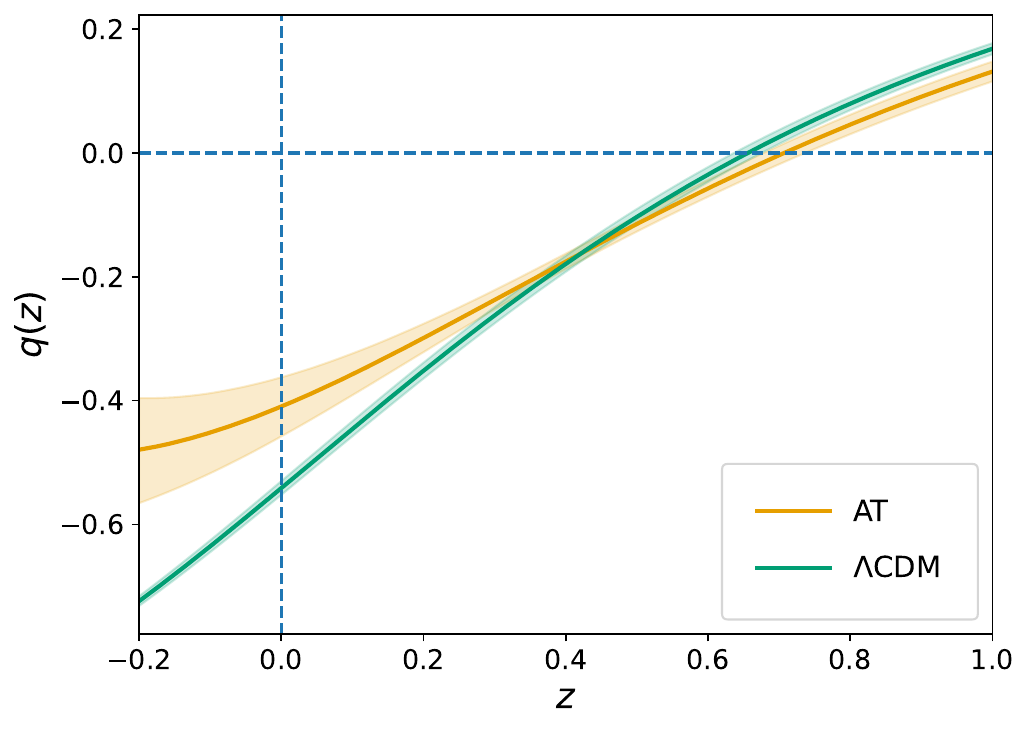}}
    \caption{The deceleration parameter, $q(z)$, reconstructed from MCMC samples (GDS dataset only) for the CPL (Fig.~\ref{7a}), TQD (Fig.~\ref{7b}), GEDE (Fig.~\ref{7c}) and AT (Fig.~\ref{7d}) DE parameterizations. The corresponding $68\%$ credible bands are shown for each parameterization. The $q(z)$ for $\Lambda$CDM, also computed from MCMC samples, is overlaid as a green curve in each panel. The redshift range $-0.2 \leq z \leq 1.0$ is chosen to illustrate the future evolution of $q(z)$.}
    \label{fig7}
\end{figure}

Now we compute the deceleration parameter $q(z) = -1 + \frac{(1+z)H'(z)}{H(z)}$ for each of the parameterizations . In Fig.~\ref{fig7}, we show the reconstructed $q(z)$ plot from the GDS dataset MCMC sample and overlay the reconstructed $q(z)$ for $\Lambda$CDM. We choose the redshift range $-0.2 \leq z \leq 1.0$ in order to show the evolution of $q(z)$ in the future. The CPL, GEDE and AT parameterizations all exhibit an accelerating expansion of the universe. However, the TQD parameterization (similar to Ref.~\cite{souza_2025}) shows a transient acceleration. The universe is expected to show deceleration starting from redshifts $z \lesssim -0.04$. Within the $68\%$ credible bands our results match Ref.~\cite{souza_2025} who showed that the deceleration phase will start at redshifts $z\sim-0.2$. The GEDE parameterization (for $\beta \geq 0$) shows a $q(z)$ curve which follows the $\Lambda$CDM curve closely. This is expected since $\beta \rightarrow 0$ for our dataset combinations and since $\Lambda$CDM is recovered for $\beta = 0$, the GEDE $q(z)$ approaches the $\Lambda$CDM curve, asymptotically. CPL and AT parameterizations follow the $\Lambda$CDM curve closely for redshifts $z \gtrsim 0.4$ but deviates at more than $1\sigma$ for lower redshifts. Nevertheless, both of them predict an accelerating universe for $z \lesssim 0.7$. 

\section{Conclusions}
\label{sec5}

In this work, we study and compare five different DE parameterizations, which also include thawing quintessence-inspired parameterizations, using a combination of non-CMB probes (BAO, SNe and GRBs) and RSD measurements. Our main results can be outlined as follows:
\begin{itemize}
    \item For the same dataset, $\Omega_{m0}$ differs between parameterizations by $ <1.2\sigma$ (Bayesian) and $ <1.8\sigma$ (frequentist). The maximum values occur for the CPL and $\Lambda$CDM frameworks with the GD dataset. In contrast, $\sigma_8$ is consistent within $0.5\sigma$ for all parameterizations and datasets, since BAO and SNe do not constrain it directly. We also find that the GRB dataset, used in this work, contribute only marginally to the constraints for all DE parameterizations. Including SNe dataset constrains the expansion history tightly and reduces the error bars on the cosmological parameters.
    \item For the CPL model, we find that $w_0$ shows a $\sim 2\sigma$ discrepancy with $w_0 = -1$ but $w_a$ values are consistent within $1\sigma$ with $w_a=0$. The EOS for CPL show a phantom crossing at $z \sim 0.5$.
    \item The TQD parameterization reduces to $\Lambda$CDM in the limit $\delta \rightarrow 0$. From Bayesian constraints, we find a $\sim 2\sigma$ discrepancy in $\delta$ values with 0, while frequentist constraints are within $1\sigma$ of 0. $\gamma$ is not bounded from above in the Bayesian methods, so we quote lower limits: $\gamma>11.52$ at $68\%$ and $\gtrsim2.83$ at $95\%$ for DS and GDS, while the frequentist analysis gives wide error bars for DS and GDS, and no constraint for GD. The TQD EOS shows a rapid decline towards $w(z) = -1$ to which it becomes asymptotic at $z \gtrsim0.3$.
    \item We restrict $\beta$ in the GEDE parameterization to attain only positive values and quote a $95\%$ upper limit (0.30 - GD, 0.11 - DS and GDS from Bayesian constraints). The EOS of GEDE shows an asymptotic curve to $w(z) = -1$ but lying below.
    \item The AT parameterization shows a discrepancy of $\sim 1.7\sigma$ (GD) and $\sim 2.6\sigma$ (DS and GDS) with $w_0 = -1$. The EOS is similar to the TQD parameterization in the sense that it asymptotically approaches $w(z) = -1$.
    \item The $q(z)$ plots, reconstructed from the MCMC samples, show an accelerating universe now and in the future. This is applicable for all parameterizations except TQD which shows an acceleration in the present but a decelerating phase in the future.
    \item Using the Bayes factor, AIC and DIC, we find that the evidence for DDE depends on both the metric and the dataset. For GD, the preferences are mostly weak or inconclusive. Including SNe (DS and GDS) strengthens the preference for extended models in AIC and DIC, while adding GRBs to DS changes the results very little. For DS and GDS, TQD is the only parameterization preferred by all three metrics, while GEDE is disfavored by all three for $\beta \geq 0$. For CPL and AT, AIC and DIC prefer the extended model (weakly for GD), but the Bayes factor weakly favors $\Lambda$CDM over CPL and is inconclusive for AT. The best-fit $\chi^2_\mathrm{min}$ improves for CPL, TQD and AT, so this difference is consistent with the Occam penalty in the Bayes factor ($\sim 4-5$ for CPL, $\sim 3-4$ for AT and $\sim 1-2$ for TQD). Hence, the preference for DDE is sensitive to prior volume, and a preference from AIC or DIC alone should be treated with caution.
\end{itemize}

In an earlier study by some of us (Ref.~\cite{barua_2025}), using DESI BAO (DR1)~\cite{desi_2026}, cosmic chronometers~\cite{moresco_2023}, quasars~\cite{cao_2022}, SNe and megamasers~\cite{pesce_2020}, $\Lambda$CDM was shown to be preferred over the CPL, Barboza-Alcaniz~\cite{BA_2008}, Jassal-Bagla-Padmanabhan~\cite{JBP_2005} and the exponential~\cite{najafi_2024}) parameterizations, and phantom-crossing was suggested in almost all of them. In this work, with DESI BAO (DR2), SNe, GRBs and RSD measurements, the thawing-quintessence-inspired parameterizations (TQD and AT) improve the fit over $\Lambda$CDM without any phantom crossing, and TQD is moderately favored by the Bayes factor for the DS and GDS dataset combination. Although the datasets differ between the two studies, both are non-CMB combinations. We find that the preference for DDE depends on the parameterization, the dataset used and the metric used for model comparison. Whether a phantom crossing is required also remains an open question.

In the absence of a compelling microphysical theory of DE, parameterizations that target different DE dynamics are useful for testing the current data. Upcoming data from the ZTF survey~\cite{ztf_2019, ztf_2025}, the Vera C. Rubin observatory~\cite{vera_2022, vera_2023} and the Nancy Grace Roman Space telescope~\cite{roman_2015} will provide larger and deeper SNe samples, which should improve the constraints on $w_0$ and $w_a$. Given the Hubble tension, it is essential to examine what non-CMB probes tell us, and not to rely only on CMB-based data.

\begin{acknowledgments}
SB would like to extend his gratitude to the University Grants Commission (UGC), Govt. of India for their continuous support through the Senior Research Fellowship, which has played a crucial role in the successful completion of our research. Computational work was supported by the National Supercomputing Mission (NSM), Government of India, through access to the ``PARAM SEVA'' facility at IIT Hyderabad. The NSM is implemented by the Centre for Development of Advanced Computing (C-DAC) with funding from the Ministry of Electronics and Information Technology (MeitY) and the Department of Science and Technology (DST).
\end{acknowledgments}


\appendix

\section{Relaxing the prior limit on the parameter $\beta$ for the GEDE parameterization.}
\label{appA}

\begin{table}[ht]
\caption{68\% Bayesian credible intervals and frequentist confidence intervals for GEDE when priors on $\beta \in [-4,2]$.}
\label{table8}
\centering

\setlength{\tabcolsep}{3pt}

\begin{tabular}{c|cc|cc|cc}
\hline
\hline

\multirow{2}{*}{\thead{Parameters}} & \multicolumn{2}{c|}{\thead{GD}} & \multicolumn{2}{c|}{\thead{DS}} & \multicolumn{2}{c}{\thead{GDS}} \\
 & \thead{Bayesian} & \thead{Frequentist} & \thead{Bayesian} & \thead{Frequentist} & \thead{Bayesian} & \thead{Frequentist} \\

\hline
\hline
&  &  &  &  &  & \\[1ex]
$H_0r_d$ [km/s] & $9979\pm140$ & $9973^{+136}_{-146}$ & $9996\pm72$ & $9991^{+76}_{-68}$ & $9995\pm72$ & $9997^{+69}_{-75}$ \\[0.7ex]
$\Omega_{m0}$ & $0.2970\pm0.0084$ & $0.2967^{+0.0081}_{-0.0087}$ & $0.2959\pm0.0082$ & $0.2948^{+0.0087}_{-0.0077}$ & $0.2961\pm0.0082$ & $0.2950^{+0.0086}_{-0.0077}$ \\[0.7ex]
$\beta$ & $-0.29^{+0.17}_{-0.22}$ & $-0.31^{+0.20}_{-0.19}$ & $-0.28^{+0.09}_{-0.11}$ & $-0.29\pm0.10$ & $-0.28^{+0.09}_{-0.11}$ & $-0.30^{+0.11}_{-0.10}$ \\[0.7ex] 
$\sigma_8$ & $0.841\pm0.037$ & $0.845^{+0.035}_{-0.038}$ & $0.840\pm0.035$ & $0.840^{+0.037}_{-0.033}$ & $0.840\pm0.035$ & $0.840^{+0.037}_{-0.033}$ \\[0.7ex]
\hline
&  &  &  &  &  &\\[0.5ex]
$\Delta \chi^2(\bar{\theta})$ & $-2.2$ & $-$ & $-6.9$ & $-$ & $-6.9$ & $-$ \\[0.7ex]
$\Delta p_D$ & $0.9$ & $-$ & $1.0$ & $-$ & $1.0$ & $-$ \\[0.7ex]
$\chi^2_\mathrm{min}$ & \multicolumn{2}{c|}{$94.9$} & \multicolumn{2}{c|}{$1614.9$} & \multicolumn{2}{c}{$1734.2$} \\[0.7ex]
$\Delta \chi^2_\mathrm{min}$ & \multicolumn{2}{c|}{$-2.3$} & \multicolumn{2}{c|}{$-6.9$} & \multicolumn{2}{c}{$-6.9$} \\ [0.7ex]
\hline
&  &  &  &  &  &\\[0.5ex]
$\ln B_{1i}$ & $1.4$ & $-$ & $-0.3$ & $-$ & $-0.3$ & $-$ \\[0.7ex]
$\Delta$DIC & $-0.4$ & $-$ & $-4.9$ & $-$ & $-4.9$ & $-$ \\[0.7ex]
$\Delta$AIC & \multicolumn{2}{c|}{$-0.3$} & \multicolumn{2}{c|}{$-4.9$} & \multicolumn{2}{c}{$-4.9$} \\[0.7ex]
\hline
\hline
\end{tabular}
\end{table}

\begin{figure}[ht]
    \centering
    \subfloat[Bayesian constraints on GEDE\label{8a}]{\includegraphics[width=0.5\textwidth,keepaspectratio]{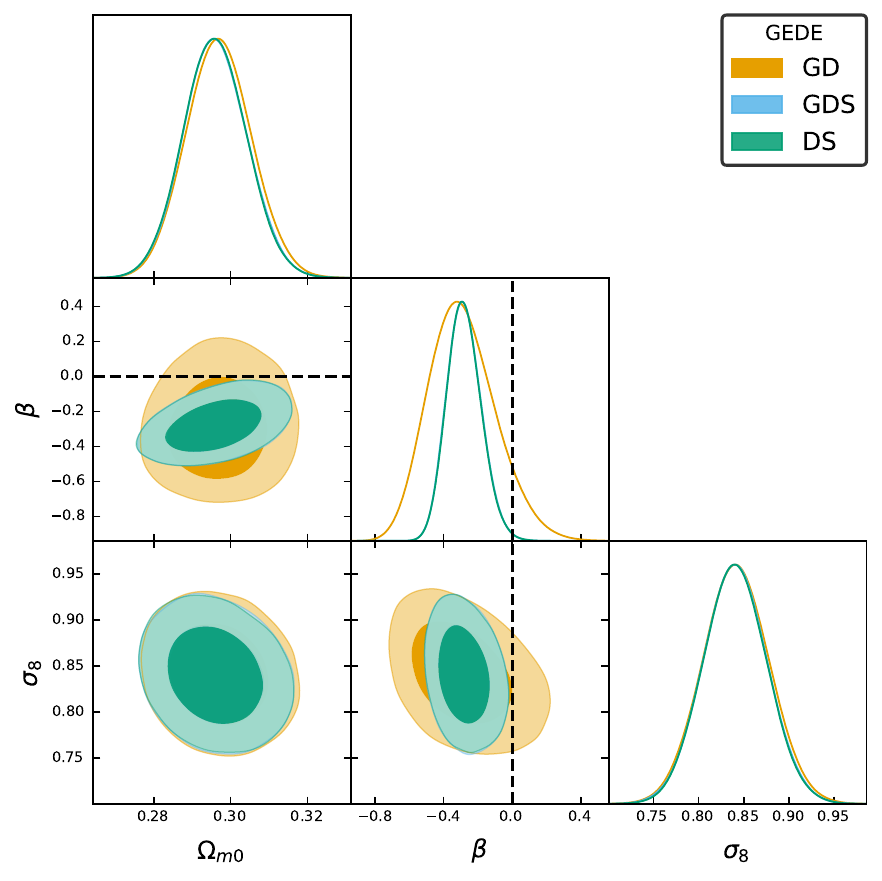}}
    \subfloat[EOS \label{8b}]{\includegraphics[width=0.5\textwidth,keepaspectratio]{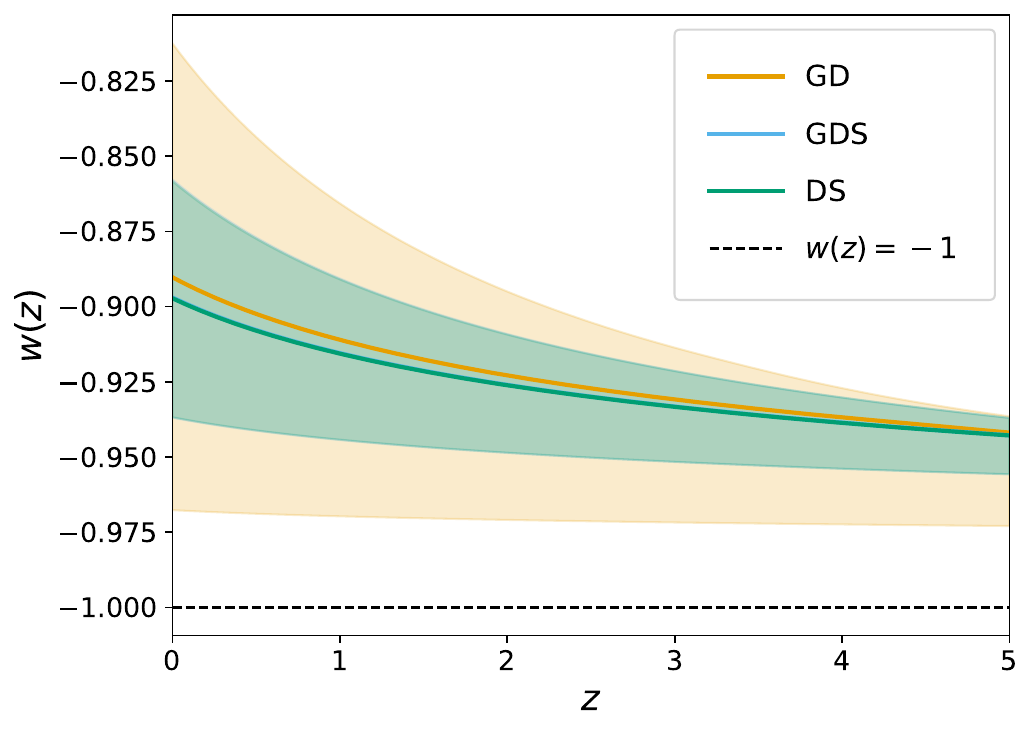}}\\
    \subfloat[$\Omega_{m0}$\label{8c}]{\includegraphics[width=0.25\textwidth,keepaspectratio]{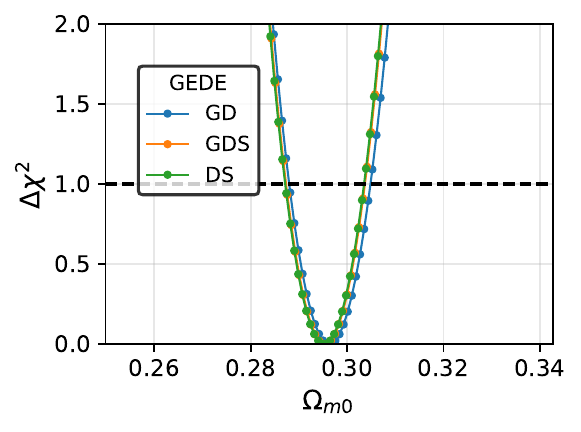}}
    \subfloat[$\beta$\label{8d}]{\includegraphics[width=0.25\textwidth,keepaspectratio]{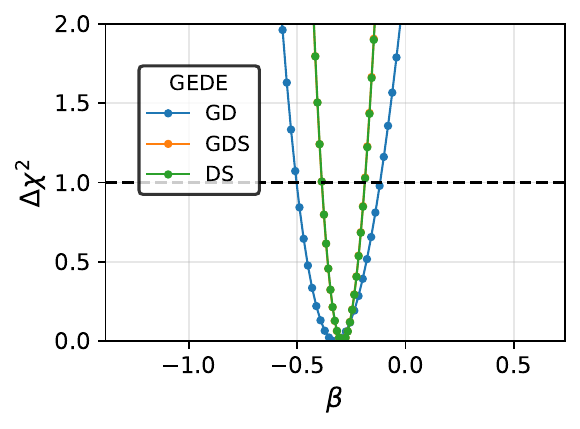}}
    \subfloat[$\sigma_8$\label{8e}]{\includegraphics[width=0.25\textwidth,keepaspectratio]{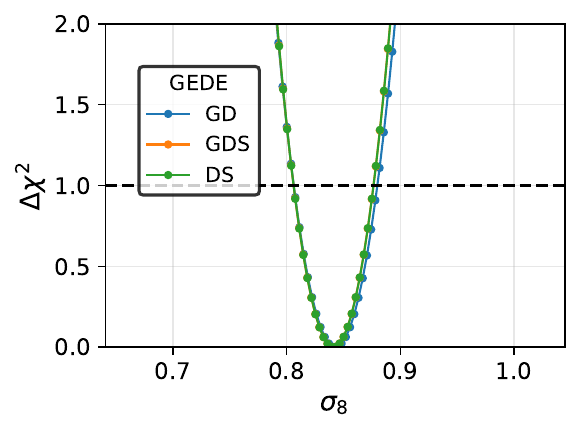}}
    \subfloat[$q(z)$\label{8f}]{\includegraphics[width=0.25\textwidth,keepaspectratio]{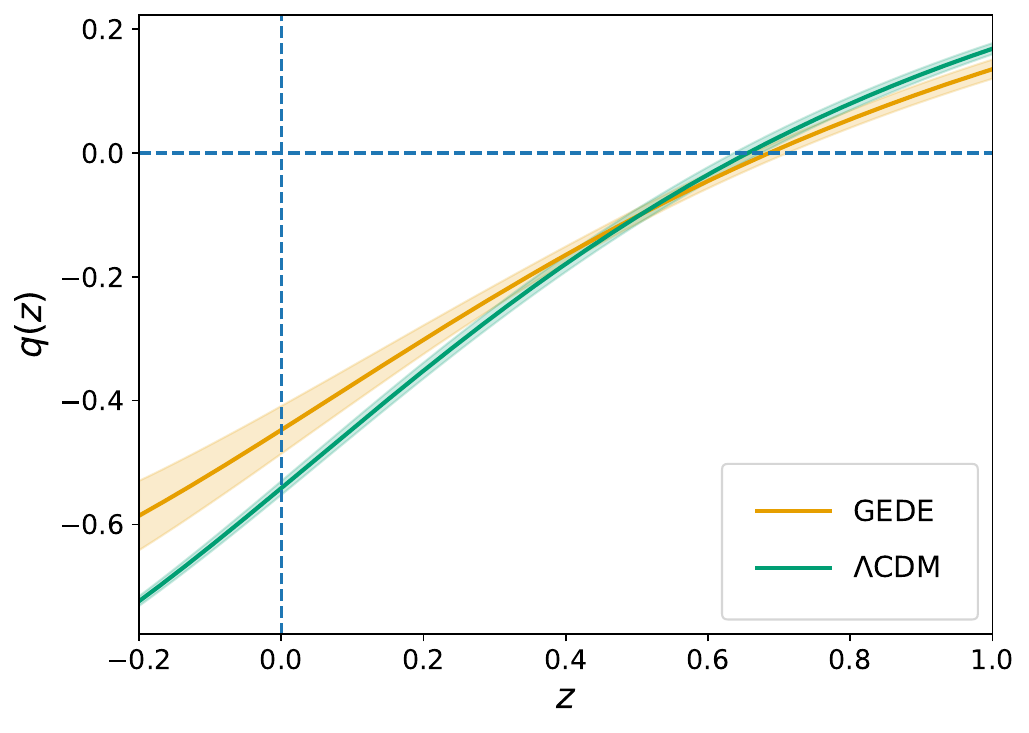}}
    \caption{Bayesian and frequentist constraints on the GEDE model with relaxed prior on $\beta$. Top row: Bayesian posterior contours (Fig.~\ref{8a}) and the corresponding EOS (Fig.~\ref{8b}); contours represent 68\% and 95\% credible regions. Bottom row: PL constraints on $\Omega_{m0}$ (Fig.~\ref{8c}), $\beta$ (Fig.~\ref{8d}) and $\sigma_8$ (Fig.~\ref{8e}). The $\Delta \chi^2 = 1$ line marks the 68\% confidence interval. Fig.~\ref{8f} shows $q(z)$, similar to Fig.~\ref{fig7}.}
    \label{fig8}
\end{figure}

\begin{figure}[ht]
    \centering
    {\includegraphics[width=0.5\textwidth,keepaspectratio]{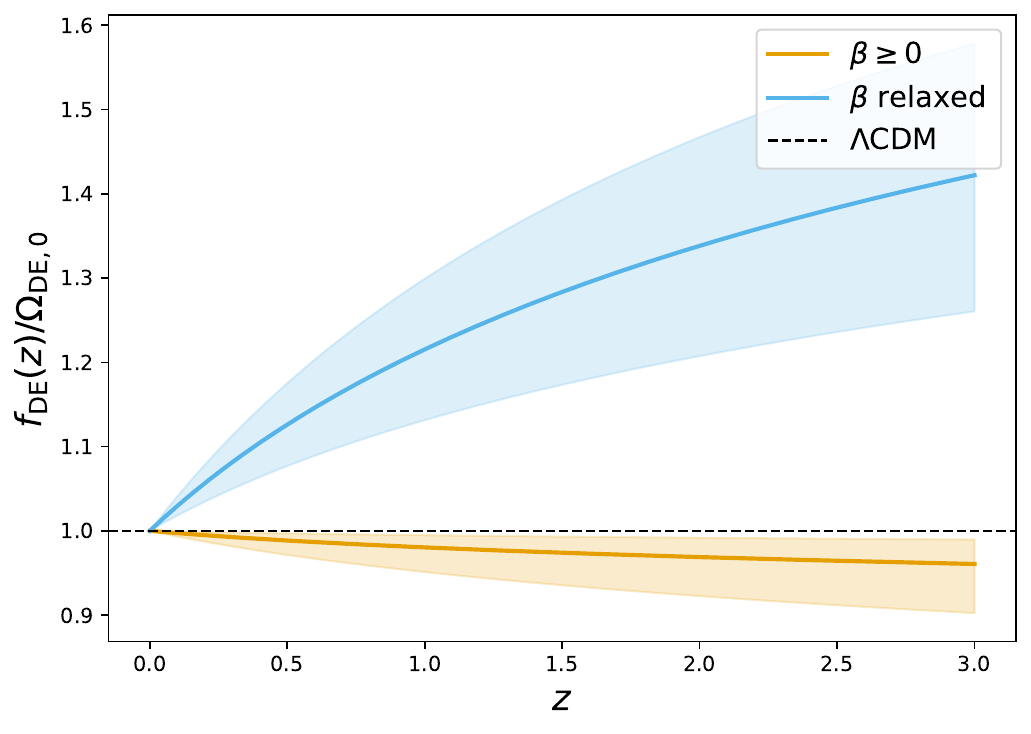}}
    \caption{Weighted median and 68\% interval of the DE density normalized to its present value,  $f_{DE}(z)/\Omega_{DE, 0}$, for GDS. $\beta \geq 0$ refers to the prior $[0, 4.34]$, $\beta $ relaxed refers to the prior $[-4, 2]$.}
    \label{fig9}
\end{figure}

In this appendix, we relax the prior on $\beta$ so that it can take negative values, $\beta \in [-4, 2]$ (Table~\ref{table8}, Fig.~\ref{fig8}). This was done in Refs.~\cite{li_2020, yang_2021} while Ref.~\cite{yang_2026} restricted $\beta$ to positive values so that the DE density stays suppressed relative to $\Lambda$CDM at high redshift. We show this in Fig.~\ref{fig9}. For $\beta \geq 0$, the orange curve lies below the $\Lambda$CDM value of 1, while the blue curve (corresponding to $\beta \in [-4, 2]$) shows a DE density that is larger than in $\Lambda$CDM. 

Relaxing the prior changes the parameter constraints in several ways. 
\begin{itemize}
    \item $\beta$: For $\beta\geq 0$, the posterior piled up against $\beta = 0$, so we could only quote a 95\% upper limit. With the relaxed prior, we obtain well-defined $68\%$ intervals in both the Bayesian and frequentist analyses. $\beta$ differs from 0 by $\sim 1.5\sigma$ (GD) and $\sim 2.8\sigma$ (DS and GDS). Hence, we notice a mild preference for $\beta < 0$. 
    \item $\Omega_{m0}$: For DS and GDS, $\Omega_{m0}$ is $1\sigma$ lower than for $\beta \in [0, 4.34]$ case.
    \item $\sigma_8$: The central value is $\sim 0.840$, compared with $0.817$ for $\beta \geq 0$. These two are consistent within $1\sigma$ because the errors are large. 
\end{itemize}

For $\beta\in[-4,2]$, the EOS lies between $-1$ and $-1 - \frac{2\beta}{3}$. For $\beta > 0$, this gives $-1 - \frac{2\beta}{3} \leq w(z) \leq -1$ (phantom), and for $\beta < 0$ it gives $-1 \leq w(z) \leq -1 + \frac{2|\beta|}{3}$ (non-phantom). This sign of $\beta$ sets which side of -1 the EOS lies on, and $w(z)$ never crosses -1 for any $\beta$. The DE density is then larger in the past than today as shown in Fig.~\ref{fig9}. Also, the $q(z)$ curves of $\Lambda$CDM and GEDE overlap for $\beta \geq 0$ (Fig.~\ref{7c}), but differ for $z \lesssim 0.5$ with the relaxed prior (Fig.~\ref{8f}).

We cannot compare our $\beta$ constraints with Refs.~\cite{yang_2021, li_2020} because the parameter definitions differ. Following Ref.~\cite{yang_2026}, we define $\beta$ as $\Delta/\ln 10$. Refs.~\cite{yang_2021, li_2020} found positive $\Delta$ using CMB data. This implies a positive $\beta$ corresponds to positive $\Delta$. Without CMB data, we therefore find the opposite sign. This suggests that the CMB data drive the preference for positive $\Delta$, although a more thorough check would be preferred. 

For the model comparison, the Bayesian evidence for the relaxed prior case is weakly in favor of $\Lambda$CDM for GD and inconclusive for DS and GDS. For $\beta \geq 0$, it was moderately in favor of $\Lambda$CDM. AIC and DIC give a weak (GD) and positive (DS and GDS) preference for GEDE over $\Lambda$CDM, in contrast with the $\beta \geq 0$ case, where $\Lambda$CDM was preferred. The Bayes factor stays inconclusive even though the fit improves because the wider prior adds an Occam penalty $\sim 3$.


\section{Studying the effect of a rapid EOS change in the AT parameterization ($p = 20.0$).}
\label{appB}

\begin{table}[ht]
\caption{68\% Bayesian credible intervals and frequentist confidence intervals for AT when $p = 20.0$.}
\label{table9}
\centering

\setlength{\tabcolsep}{3pt}

\begin{tabular}{c|cc|cc|cc}
\hline
\hline

\multirow{2}{*}{\thead{Parameters}} & \multicolumn{2}{c|}{\thead{GD}} & \multicolumn{2}{c|}{\thead{DS}} & \multicolumn{2}{c}{\thead{GDS}} \\
 & \thead{Bayesian} & \thead{Frequentist} & \thead{Bayesian} & \thead{Frequentist} & \thead{Bayesian} & \thead{Frequentist} \\

\hline
\hline
&  &  &  &  &  & \\[1ex]
$H_0r_d$ [km/s] & $9822\pm210$ & $9824^{+200}_{-210}$ & $9979\pm71$ & $9984^{+68}_{-75}$ & $9979\pm71$ & $9976^{+75}_{-68}$ \\[0.7ex]
$\Omega_{m0}$ & $0.3130\pm0.0120$ & $0.3121^{+0.0130}_{-0.0116}$ & $0.3050\pm0.0074$ & $0.3044^{+0.0078}_{-0.0069}$ & $0.3052\pm0.0074$ & $0.3053^{+0.0071}_{-0.0077}$ \\[0.7ex]
$w_0$ & $-0.10\pm0.53$ & $-0.08^{+0.52}_{-0.53}$ & $-0.51\pm0.16$ & $-0.50^{+0.15}_{-0.17}$ & $-0.51\pm0.16$ & $-0.52^{+0.17}_{-0.15}$ \\[0.7ex] 
$\sigma_8$ & $0.837\pm0.035$ & $0.836^{+0.037}_{-0.033}$ & $0.831\pm0.034$ & $0.829^{+0.036}_{-0.032}$ & $0.831\pm0.034$ & $0.829^{+0.036}_{-0.032}$ \\[0.7ex]
\hline
&  &  &  &  &  &\\[0.5ex]
$\Delta \chi^2(\bar{\theta})$ & $-2.9$ & $-$ & $-9.0$ & $-$ & $-9.0$ & $-$ \\[0.7ex]
$\Delta p_D$ & $1.0$ & $-$ & $1.0$ & $-$ & $1.0$ & $-$ \\[0.7ex]
$\chi^2_\mathrm{min}$ & \multicolumn{2}{c|}{$94.3$} & \multicolumn{2}{c|}{$1612.8$} & \multicolumn{2}{c}{$1732.1$} \\[0.7ex]
$\Delta \chi^2_\mathrm{min}$ & \multicolumn{2}{c|}{$-2.9$} & \multicolumn{2}{c|}{$-9.0$} & \multicolumn{2}{c}{$-9.0$} \\ [0.7ex]
\hline
&  &  &  &  &  &\\[0.5ex]
$\ln B_{1i}$ & $-0.1$ & $-$ & $-2.0$ & $-$ & $-2.0$ & $-$ \\[0.7ex]
$\Delta$DIC & $-0.9$ & $-$ & $-7.0$ & $-$ & $-7.0$ & $-$ \\[0.7ex]
$\Delta$AIC & \multicolumn{2}{c|}{$-0.9$} & \multicolumn{2}{c|}{$-7.0$} & \multicolumn{2}{c}{$-7.0$} \\[0.7ex]
\hline
\hline
\end{tabular}
\end{table}

\begin{figure}[ht]
    \centering
    \subfloat[Bayesian constraints on AT\label{10a}]{\includegraphics[width=0.5\textwidth,keepaspectratio]{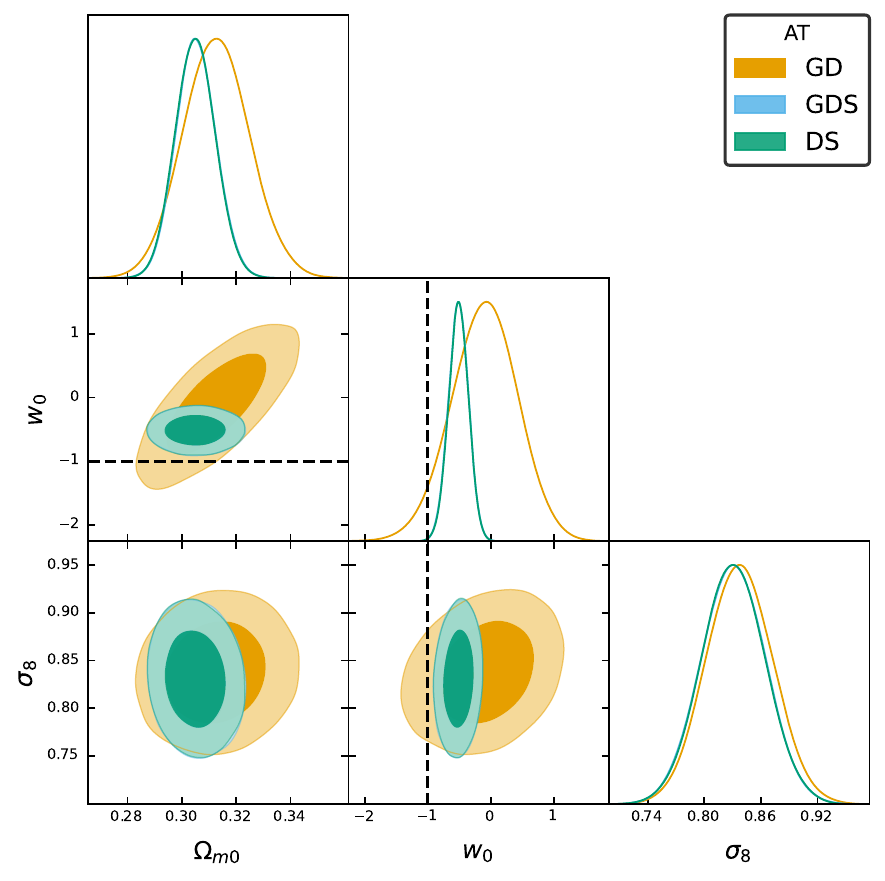}}
    \subfloat[EOS \label{10b}]{\includegraphics[width=0.5\textwidth,keepaspectratio]{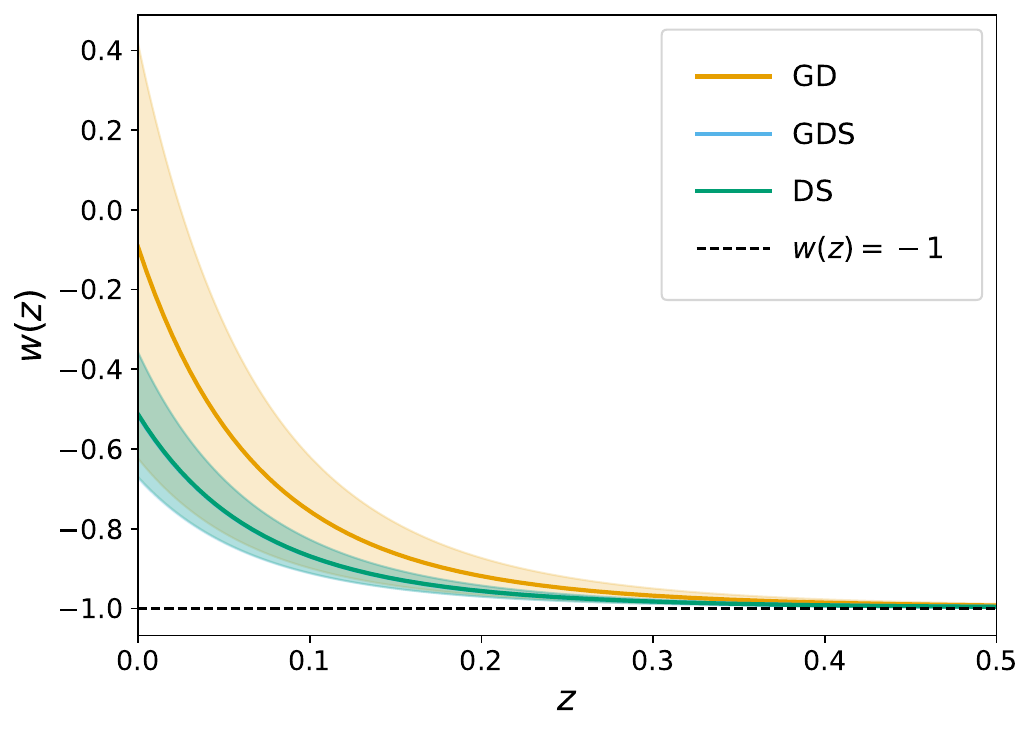}}\\
    \subfloat[$\Omega_{m0}$\label{10c}]{\includegraphics[width=0.25\textwidth,keepaspectratio]{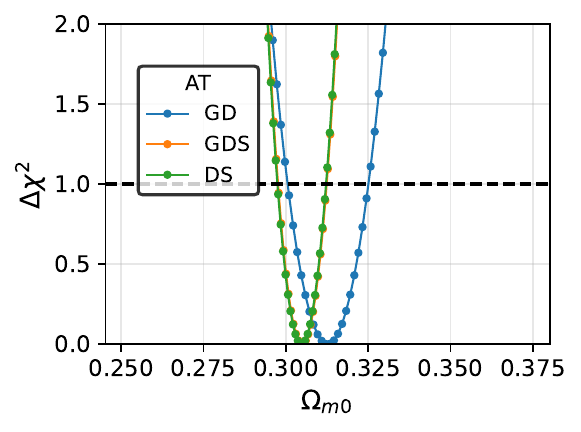}}
    \subfloat[$w_0$\label{10d}]{\includegraphics[width=0.25\textwidth,keepaspectratio]{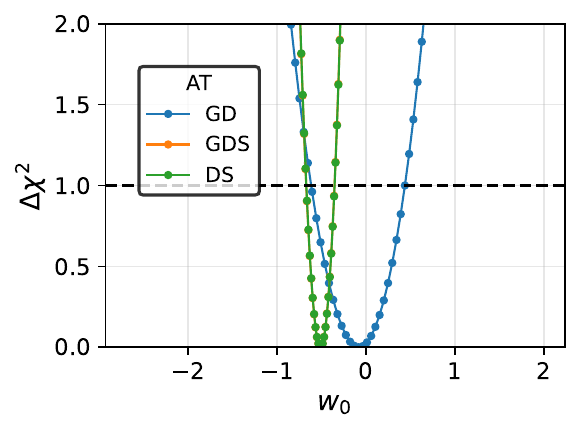}}
    \subfloat[$\sigma_8$\label{10e}]{\includegraphics[width=0.25\textwidth,keepaspectratio]{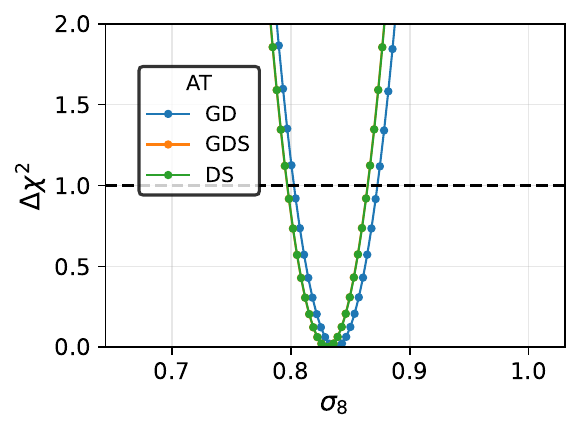}}
    \subfloat[$q(z)$\label{10f}]{\includegraphics[width=0.25\textwidth,keepaspectratio]{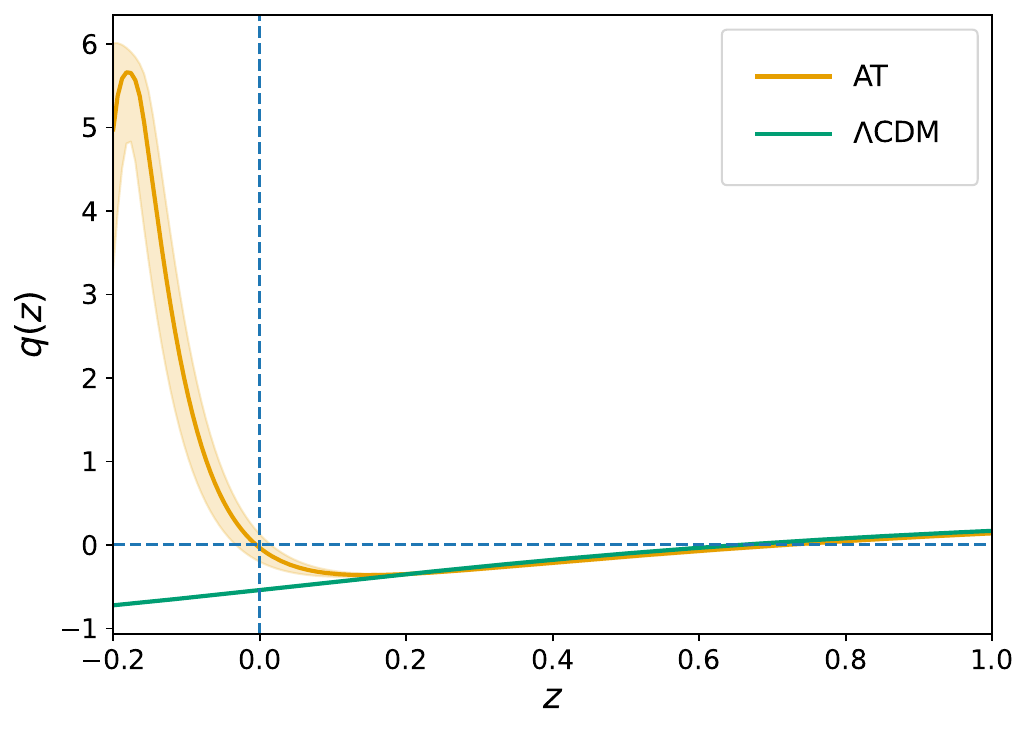}}
    \caption{Bayesian and frequentist constraints on the AT model with p = 20. Top row: Bayesian posterior contours (Fig.~\ref{10a}) and the corresponding EOS (Fig.~\ref{10b}); contours represent 68\% and 95\% credible regions. Bottom row: PL constraints on $\Omega_{m0}$ (Fig.~\ref{10c}), $w_0$ (Fig.~\ref{10d}) and $\sigma_8$ (Fig.~\ref{10e}). The $\Delta \chi^2 = 1$ line marks the 68\% confidence interval. Fig.~\ref{10f} shows $q(z)$, similar to Fig.~\ref{fig7}.}
    \label{fig10}
\end{figure}

In this appendix, we study how a rapid evolution of $w(z)$ in AT, controlled by $p$, affects parameter constraints (Table~\ref{table9}, Fig.~\ref{fig10}). Ref.~\cite{lodha_2025} kept $p$ free and found that it hit the prior boundary at 30. We instead adopt a conservative approach and fix $p = 20$, which stays away from this boundary. Compared to $p=20$, $w_0$ for $p=1$ is lower by $\sim 1.3\sigma$ (GD) and $\sim 2\sigma$ (DS and GDS). The discrepancy of $w_0$ with $-1$ is $\sim 1.7 \sigma$ (GD) and $\sim 2.6\sigma$ (DS and GDS) for $p=1$, while for $p=20$ it is $\sim 3 \sigma$ for DS and GDS and $1.7 \sigma$ for GD. Hence, the rapid evolution of $w(z)$ increases the deviation of $w_0$ from $-1$ only slightly (the error bars on $w_0$ increase when $p$ increases). The $q(z)$ curve shows that the universe is accelerating today and decelerates again in the very near future (Fig.~\ref{10f}).

For model comparison, the Bayes factor weakly prefers AT with $p=20$ over $\Lambda$CDM for DS and GDS and is inconclusive for GD. For $p=1$, it was inconclusive for DS and GDS and weakly in favor of $\Lambda$CDM for GD. AIC and DIC strongly favor AT with $p=20$ for DS and GDS, compared to a positive preference for $p=1$. The Occam penalty is $\sim 1.4$ (GD) and $\sim 2.5$ (DS and GDS), smaller than $p=1$. 

\bibliography{references}

\end{document}